\documentclass[aps,prx,amssymb,twocolumn,amsmath,superscriptaddress,longbibliography,reprint,10pt,tightenlines,dvipdfmx]{revtex4-2}

\usepackage{appendix}
\usepackage{amsmath,amssymb}
\usepackage{mathtools}
\mathtoolsset{showonlyrefs,showmanualtags}
\usepackage{physics}
\usepackage{tikz}
\usepackage{bm}
\usepackage{braket}

\usepackage{graphicx}
\usepackage{here}
\usepackage[breaklinks,  hyperfootnotes=true, pdfpagelabels, bookmarks, pageanchor]{hyperref}
\hypersetup{
colorlinks=true, linktocpage=true, pdfstartpage=1, pdfstartview=FitH, pdfborder={0 0 0}, breaklinks=true, pdfpagemode=UseNone, pageanchor=true, pdfpagemode=UseOutlines, plainpages=false, bookmarksnumbered, bookmarksopen=true, bookmarksopenlevel=1, hypertexnames=true, pdfhighlight=/O, urlcolor=blue, linkcolor=blue, citecolor=blue}

\newcommand{\opt}{\mathrm{opt}}
\newcommand{\num}{\mathrm{num}}
\newcommand{\ri}{\mathrm{(i)}}
\newcommand{\rii}{\mathrm{(ii)}}
\newcommand{\riii}{\mathrm{(iii)}}
\newcommand{\dbra}[1]{\ensuremath{\langle\!\langle#1|}}
\newcommand{\dket}[1]{\ensuremath{|#1\rangle\!\rangle}}
\newcommand{\dbraket}[2]{\ensuremath{\langle\!\langle #1 | #2 \rangle\!\rangle}}

\begin{document}
\title{Analytical derivation and extension of the anti-Kibble-Zurek scaling in the transverse field Ising model}

\author{Kaito Iwamura}
\affiliation{Department of Physics, Waseda University, Tokyo 169-8555, Japan}

\author{Takayuki Suzuki}
\affiliation{National Institute of Information and Communications Technology, Nukui-Kitamachi 4-2-1, Koganei, Tokyo 184-8795, Japan}

\begin{abstract}
A defect density which quantifies the deviation from the spin ground state characterizes non-equilibrium dynamics during phase transitions. The widely recognized Kibble--Zurek scaling predicts how the defect density evolves during phase transitions. However, it can be perturbed by a noise, leading to the anti-Kibble--Zurek scaling. In this research, we analytically investigate the effect of Gaussian white noise on the transition probabilities of the Landau--Zener model. We apply this analysis to the one-dimensional transverse field Ising model and obtain an analytical approximate solution of the defect density. Our analysis reveals that when the introduced noise is small, the model follows the previously known anti-Kibble--Zurek scaling. Conversely, when the noise increases, the scaling can be obtained by using the adiabatic approximation. This result indicates that deriving the anti-Kibble--Zurek scaling does not require solving differential equations, instead, it can be achieved simply by applying the adiabatic approximation. Furthermore, we identify the parameter that minimizes the defect density based on the new scaling, which allows us to verify how effective the already known scaling of the optimized parameter is.
\end{abstract}

\maketitle

\section{Introduction}\label{sec:intro}
Dynamical phase transitions, where the properties change in time-evolving systems, are fascinating phenomena. Advancements in quantum technology are making it increasingly possible to control quantum dynamics in various physical systems, thereby making the study of these phenomena increasingly important. 

One related area of phase transitions is adiabatic quantum computation~\cite{farhi2001quantum, santoro2002theory}. This approach involves a quantum algorithm that guides the evolution of the system over time, ensuring that it moves towards the local ground states of the Hamiltonian of interest. Errors in this process can arise from excitations caused by non-adiabatic effects during dynamical evolution. This issue can be mitigated by evolving the system more slowly over time. However, if the computation time is too long, noise from the external environment can lead to additional excitations and errors. Therefore, understanding the role of dissipation in the adiabatic processes in the many-body systems is crucial for improving the performance of adiabatic quantum computation.

The Kibble–-Zurek mechanism~\cite{zurek2005dynamics,
damski2005simplest,
dziarmaga2005dynamics,
polkovnikov2005universal} (KZM) provides valuable theoretical insights into adiabatic quantum computation. The KZM was originally proposed within cosmology~\cite{kibble1976topology,kibble1980some} and later applied to condensed matter systems~\cite{zurek1985cosmological,
zurek1993cosmic,zurek1996cosmological}. The KZM gives a theoretical framework for understanding the formation of defects in systems undergoing quantum phase transitions.

The transverse field XY model and the transverse field Ising model induce quantum phase transitions. These are systems, where the spins interacting with each other are subjected to an external control field. The KZM allows us to estimate the defect density, which represents deviations from the ground state, in the adiabatic regime~\cite{zurek2005dynamics,damski2005simplest,dziarmaga2005dynamics,polkovnikov2005universal,
mukherjee2007quenching}. The time dependence of the Hamiltonian due to the application of an external control field is characterized by the parameter of the driving speed $v$. In the context of quantum control, it is desirable to minimize defects. According to the KZM, as the $v$ decreases, the defect density decreases. The region where $v$ is small is referred to as the adiabatic region. 

Recent studies, however, have pointed out an interesting mechanism called the anti-Kibble--Zurek mechanism (anti-KZM)~\cite{griffin2012scaling,gao2017anti,dutta2016anti,puebla2020universal,singh2023driven,singh2021driven}. The anti-KZM argues that, in noise-induced systems, defects become larger when the driving speed $v$ is small, which is contrary to the predictions of the KZM. In some experiments, the anti-KZM has been observed by deriving scaling in the transverse Ising model~\cite{polkovnikov2011colloquium,cui2016experimental,ai2021experimental,wang2014quantum,gong2016simulating} or with other systems~\cite{ulm2013observation,clark2016universal,chen2011quantum,pyka2013topological,del2013causality,braun2015emergence,anquez2016quantum}.

A previous research simulates the defect density in the transverse field XY model with Gaussian white noise~\cite{gao2017anti}. The transverse field XY model is a kind of quantum spin system in which an external control field is uniformly applied to the spin-chain, where spins interact with the nearest-neighbor spins. The Hamiltonian is given by
\begin{align}
&\quad {H}_{\text{XY}}(t)\\
&=-\sum^{N}_{j=1}\qty(J_x{\sigma}^x_{j}{\sigma}^x_{j+1}+J_y{\sigma}^y_{j}{\sigma}^y_{j+1}+h{\sigma}^z_{j})\\
&=-\frac{K}{2}\sum^{N}_{j=1}\qty((1+\phi){\sigma}^x_{j}{\sigma}^x_{j+1}+(1-\phi){\sigma}^y_{j}{\sigma}^y_{j+1})-h\sum^{N}_{j=1}{\sigma}^z_{j}.
\end{align}
Here, $J_x$, $J_y$, and $K=J_x+J_y$ are the the nearest-neighbor interaction constants, $\phi=\frac{J_x-J_y}{K}$ is the anisotropy, and $h$ denotes the external control field. ${\sigma}^a_{j}$ ($a=x,y,z$) are Pauli matrices, which acts on the $j$-th spin. The number of spins $N$ is assumed to be sufficiently large. The system undergoes a quantum phase transition when $h$, $J_x$, $J_y$, or $\phi$ evolve over time and cross the critical point. The time-dependence of ${H}_{\text{XY}}$ is introduced through either $h(t)=vt+\gamma(t)$, $J_x(t)=vt+\gamma(t)$, or $\phi(t)=vt+\gamma(t)$. Here, $v$ is the constant driving speed, and $\gamma(t)$ denotes Gaussian white noise with the zero mean and the second moment
\begin{align}
\overline{\gamma(t)}&=0,\\ \overline{\gamma(t)\gamma(t')}&=W^2\delta(t-t').
\label{eq:white_noise}
\end{align}
The scaling of the defect density $n$ for small $W^2$ in the above three cases is represented as~\cite{gao2017anti}
\begin{align}
n\simeq cv^{\beta}+rv^{-1}.
\label{eq:scaling}
\end{align}
Here, $c$, $r$, and $\beta$ are constants determined by the choice of the system and $r$ depends on $W^2$. The first term of Eq.~\eqref{eq:scaling} represents the well-known term identified in the KZM and the second term which is proportional to $v^{-1}$ emerges due to the introduction of noise. This is a numerically established result and have not analytically derived.

In a numerical research on the noise-induced transverse field Ising model with Gaussian white noise~\cite{dutta2016anti}, under the following setup $J_x(t)=vt+\gamma(t),\ J_y=0,\ h(t)=J-vt-\gamma(t),\ t\in[0,J/v]$. The similar anti-KZM scaling as in Eq.~\eqref{eq:scaling} has been confirmed. Note that the scaling of $v^{-1}$ in Eq.~\eqref{eq:scaling} leads to a divergence for small $v$. As stated in~\cite{dutta2016anti}, however, the Kayanuma formula~\cite{kayanuma1984nonadiabatic} provides an asymptotic expression, in the small $v$ regime,
\begin{align}
n^{\text{Kayanuma}} \simeq \frac{1}{2} - \frac{\sqrt{v}}{4\pi J},
\end{align}
which does not diverge. This formula is derived under the approximation that the noise is sufficiently large. It converges asymptotically to 1/2 as $v$ approaches zero, which is consistent with the numerical simulation. However, if noise is small, this approximate solution is not valid in most regions. 

The anti-KZM suggests that there exists a specific non-zero optimized driving speed, denoted by $v_{\opt}$, which minimizes the defect density. The optimized driving speed is considered to be the ideal parameter in quantum control. Both numerical calculations~\cite{gao2017anti,dutta2016anti} examine the relationship between the optimized driving speed $v_{\opt}$ and the parameter $W^2$, which represents the magnitude of the noise. They confirm that in the small $W^2$ regime, $v_{\opt}$ scales as $v_{\opt} \propto W^{4/3}$.

Note that, to obtain the anti-KZM scaling, one method involves calculating the defect density using Gaussian colored noise and then taking the white-noise limit~\cite{singh2021driven,singh2023driven}. Gaussian colored noise $\tilde{\gamma}(t)$ satisfies $\overline{\tilde{\gamma}(t)}=0$ and $\overline{\tilde{\gamma}(t)\tilde{\gamma}(t')}=\tilde{W}^2 e^{-\Gamma|t-t'|}$. White noise can serve as a simpler approximation for colored noise when the energy spectrum of the colored noise decays slowly. Consequently, the white noise tends to provide higher energy for these excitations compared to colored noise. Nevertheless, both types of noise are expected to exhibit qualitatively similar anti-KZM behavior. Thus, using white noise is sufficient to capture the essential anti-KZM scaling.

Other studies have explored systems with various perturbations, such as a dissipative thermal bath~\cite{patane2008adiabatic,patane2009adiabatic,yin2014nonequilibrium,nalbach2015quantum,keck2017dissipation,arceci2018optimal,nigro2019competing,rossini2019scaling,rossini2020dynamic}, a recycling term~\cite{dora2019kibble,gulacsi2021defect}, and a time-dependent simple harmonic oscillator~\cite{mukherjee2009effects,mullen1989time,kayanuma2000landau,suzuki2023kibble}.

This research aims to analytically derive the defect density and reproduce the anti-KZM in the one-dimensional transverse field Ising model with Gaussian white noise in the transverse field term. Introducing this noise simulates imperfections in system control that arise from the inability to apply a perfectly uniform static control field. In addition, we derive a more precise scaling than the previously proposed the anti-KZM scaling in unexplored regions. For this purpose, we first perform an analytical derivation of the transition probabilities from the ground state in the Landau--Zener model with Gaussian white noise. Second, by applying approximate solutions of the Landau--Zener model to the transverse field Ising model. Finally, we analytically determine the optimized driving speed $v_{\opt}$. We then compare it with the scaling $v_{\opt} \propto W^{4/3}$ observed in the numerical calculations~\cite{gao2017anti,dutta2016anti}.

In Sec.~\ref{sec:SETUP}, we derive the transition probability of the Landau--Zener model with Gaussian white noise. In Sec.~\ref{sec:ising}, we apply the formula obtained in Sec.~\ref{sec:SETUP} to the transverse field Ising model with Gaussian white noise in the adiabatic limit to determine the scaling of the defect density. Section~\ref{sec:vopt} is dedicated to a discussion on the parameters that minimize the defect density. Section~\ref{sec:conclusion} is devoted to the conclusion.

\section{The Landau--Zener model with Gaussian white noise}\label{sec:SETUP}
Consider the following two-level Hamiltonian system, in the natural units,
\begin{align}
\label{eq:LZ_noise}
{H}_{\text{LZ+noise}}(t)= {H}_{\text{LZ}}(t)+{H}_{\text{noise}}(t),
\end{align}
\begin{align}
{H}_{\text{LZ}}(t)= \frac{vt}{2}{\sigma}^z+{J}{\sigma}^x,\quad {H}_{\text{noise}}(t)=\gamma(t){\sigma}^z.
\end{align}
Here, the term ${H}_{\text{LZ}}$ represents a Landau--Zener model~\cite{landau1932theorie,zener1932non,stuckelberg1932theorie,majorana1932atoms,ivakhnenko2023nonadiabatic}. The model describes a quantum two-level system with a linearly time-dependent energy difference, employed to analytically calculate transition probabilities from the ground state to the excited state. $J$ denotes the interaction constant of an energy dimension and $v$ denotes the driving speed with a dimension of energy squared. The time-evolution operator is obtained with the parabolic cylinder functions. $\gamma(t)$ in ${H}_{\text{noise}}$ represents Gaussian white noise satisfying Eq.~\eqref{eq:white_noise}, where $W^2$ has the dimension of energy squared multiplied by time. The quantities with an overline indicate that they are ensemble-averaged.

The expectation value of a time-independent observable $P$ at an arbitrary time $t$ can be expressed as
\begin{align}
\ev{\overline{P(t)}}=\Tr[\overline{\rho(t)}P]=\Tr[\overline{U(t,t_i)\rho(t_i)U^{\dagger}(t,t_i)}P],
\label{eq:ev_P}
\end{align}
where the density matrix $\rho(t)$ and the time-evolution operator $U(t,t_i)$ include contributions of noise. The time $t_i\to-\infty$ indicates an initial time and $U(t_i,t_i)=1$. The noise-averaged density matrix $\overline{\rho(t)}$ follows the master equation
\begin{align}
\frac{d}{dt}\overline{\rho(t)}=-i\qty[{H}_{\text{LZ}}(t),\overline{\rho(t)}]+\frac{1}{2}W^2\qty[\qty[{\sigma}^z,\overline{\rho(t)}],{\sigma}^z].
\end{align}
This equation is derived in~\cite{dutta2016anti} by Novikov's theorem~\cite{novikov1965functionals} and is derived also in ~\cite{pichler2013heating,rahmani2015dynamics}. Employing the dimensionless parameters: the time parameter $\tau=\sqrt{v}t$, the adiabatic parameter $\kappa={J}^2/v$, and the noise strength parameter $\lambda=W^2/J$, the master equation is expressed as
\begin{align}
\frac{d}{d\tau}\overline{\rho(\tau)}
=&-i\qty[{H}_{\text{LZ}}(\tau),\overline{\rho(\tau)}]\\
&+\frac{\lambda}{2}\sqrt{\kappa}\qty[\qty[{\sigma}^z,\overline{\rho(\tau)}],{\sigma}^z] ,\label{eq:mastereq}
\end{align}
\begin{align}
{H}_{\text{LZ}}(\tau)= \frac{\tau}{2}{\sigma}^z+\sqrt{\kappa}{\sigma}^x.
\end{align}

At the initial time $\tau_i$ chosen to be the remote past, the initial state is set as the ground state $\rho(\tau_i)=\ket{\uparrow}\bra{\uparrow}$. Here, ${\sigma}^z \ket{\uparrow}=\ket{\uparrow}$, ${\sigma}^z \ket{\downarrow}=-\ket{\downarrow}$. If the state $\ket{\uparrow}$ is evolved adiabatically without noise, it transitions to the state $\ket{\downarrow}$ at the final time. Therefore, the excitation state at the final time is $\ket{\uparrow}$, and the transition probability is obtained by inserting $P=\ket{\uparrow}\bra{\uparrow}$ into Eq.~\eqref{eq:ev_P}. The first term on the right-hand side of Eq.~\eqref{eq:mastereq} represents the unitary term. If only this term is present, the analytical solution for the density matrix $\overline{\rho(\tau)}$ is obtained with the parabolic cylinder functions, which yields a well-known result $\ev{\overline{P(\infty)}}= e^{-2\pi\kappa}$. As the parameter $\kappa$ increases, the adiabaticity increases and the transition probability decreases. The second term on the right-hand side in Eq.~\eqref{eq:mastereq}, the dissipation term, accounts for the interaction with noise. The presence of this term leads to an increase in the transition probability as $\kappa$ increases. Note that if we consider the infinite time span between remote past and future, a finite shift of the time $\tau$ of ${H}_{\text{LZ}}$ does not affect the transition probability. This is because the master equation~\eqref{eq:mastereq} explicitly depends on time only through ${H}_{\text{LZ}}$.

In this study, we consider the case where the noise is sufficiently weak, i.e., $\lambda\ll1$. We divide $\kappa$ into three distinct regions: (i) the small region ($0<\kappa\ll 1$), (ii) the intermediate region ($1\ll \kappa\ll 1/\lambda$), and (iii) the large region ($1/\lambda\ll\kappa$). In the regions (i) and (ii), we employ a perturbation up to the first order in $\lambda$ in Sec.~\ref{subsec:1st}. On the other hand, in the regions (ii) and (iii), we employ the adiabatic approximation in Sec.~\ref{subsec:adiabatic}. By combinating these approximations, we obtain a globally valid estimation of the transition probability in Sec.~\ref{subsec:all}.
\begin{figure}[H]
\centering
\includegraphics[width=75mm]{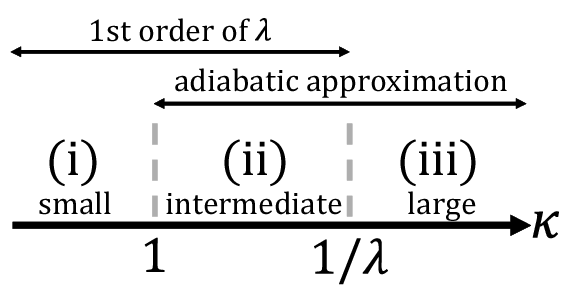}
\caption{A schematic diagram illustrating the domains of $\kappa$: (i), (ii), and (iii). Over each region, the corresponding approximation methods are indicated.}
\label{figure/region}
\end{figure}

Note that while previous studies~\cite{kayanuma1984nonadiabatic,pokrovsky2003fast,kenmoe2013effects,ashhab2006decoherence,chen2024noisy} have considered adding Gaussian colored noise $\tilde{\gamma}(t)$ to the Landau-Zener model, none have provided an analytical solution for transition probability by taking the white-noise limit to match our setting.

\subsection{First-order perturbation for noise}\label{subsec:1st}
\begin{figure}[H]
\centering
\includegraphics[width=85mm]{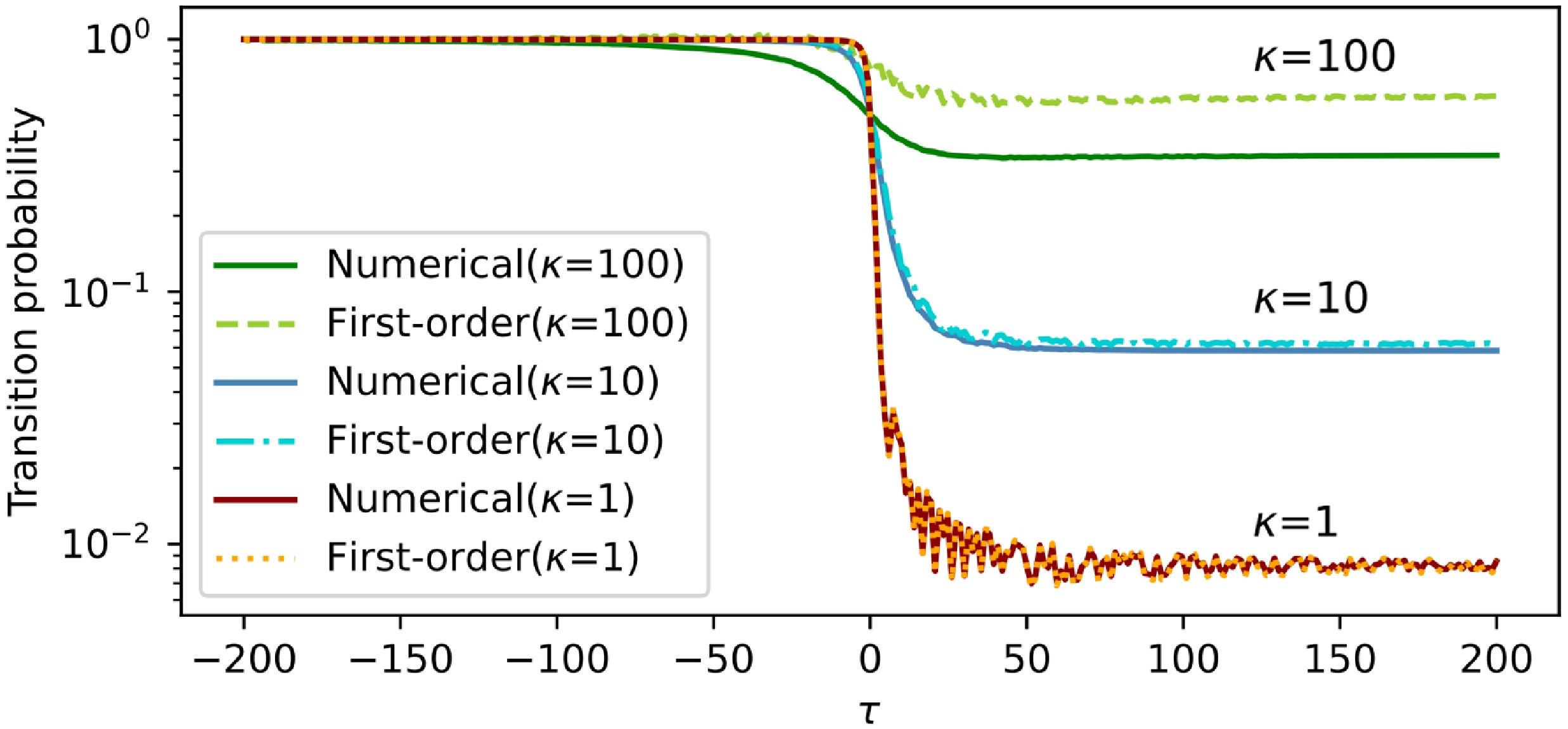}
\caption{Comparison between numerical results obtained by solving the master equation~\eqref{eq:mastereq_int} and Eq.~\eqref{eq:approx_1st_numerical}, which is first-order approximation in $\lambda$. The solid lines (Numerical) represent the numerical results. The dashed lines (first order) represent the first-order approximation. Both simulations cover a time interval from $\tau_i=-200$ to $\tau_f=200$, with $\lambda$ set to be $10^{-3}$ and $\kappa$ is set to be 100, 10, and 1.}
\label{figure/rho_time_compare}
\end{figure}

In this subsection, we consider (i) small region ($\kappa\ll1$) and (ii) intermediate region ($1\ll\kappa\ll 1/\lambda$). In these regions, we approximate the solution up to the first order in $\lambda$. If we denote $\ket{\uparrow}=(1\ 0)^T$ and $\ket{\downarrow}=(0\ 1)^T$, the time-evolution operator $U_{\text{LZ}}$ for ${H}_{\text{LZ}}$ is expressed as
\begin{align}
\label{eq:unitary}
U_{\text{LZ}}(\tau,\tau_i)=&
\begin{pmatrix}
f(\tau,\tau_i)&-g^\ast(\tau,\tau_i)\\
g(\tau,\tau_i)&f^\ast(\tau,\tau_i)
\end{pmatrix},
\end{align}
\begin{align}
f\left(\tau, \tau_i\right)=&e^{-\pi \kappa / 2} D_{i \kappa}\left(e^{-\frac{\pi}{4} i} \tau_i\right) D_{-i \kappa}\left(e^{\frac{\pi}{4} i} \tau\right) \\
&{\hspace{-0.05in}}+e^{-\pi \kappa / 2} \sqrt{\kappa} D_{-i \kappa-1}\left(e^{\frac{\pi}{4} i} \tau_i\right) \sqrt{\kappa} D_{i \kappa-1}\left(e^{-\frac{\pi}{4} i} \tau\right),\\
g\left(\tau, \tau_i\right)=&e^{-\pi \kappa / 2} e^{\frac{\pi}{4} i} D_{i \kappa}\left(e^{-\frac{\pi}{4} i} \tau_i\right) \sqrt{\kappa} D_{-i \kappa-1}\left(e^{\frac{\pi}{4} i} \tau\right)\\
&{\hspace{-0.05in}}-e^{-\pi \kappa / 2} e^{\frac{\pi}{4} i} \sqrt{\kappa} D_{-i \kappa-1}\left(e^{\frac{\pi}{4} i} \tau_i\right) D_{i \kappa}\left(e^{-\frac{\pi}{4} i} \tau\right).
\end{align}
Here, $D_{\nu}(z)$ is the parabolic cylinder function~\cite{gradshteyn2014table}. From the relation, valid for any 
$\tau\in \mathbb{R}$,
\begin{align}
e^{-\pi \kappa / 2}\left( \left|D_{i \kappa}\left(e^{-\frac{\pi}{4} i} \tau\right)\right|^2+\left|\sqrt{\kappa} D_{i \kappa-1}\left(e^{-\frac{\pi}{4} i} \tau\right)\right|^2\right)=1,
\end{align}
the initial condition $U(\tau_i,\tau_i)=1$ is satisfied. Moving to the interaction picture
\begin{align}
\tilde{\rho}(\tau)=&U^{\dagger}_{\text{LZ}}(\tau,\tau_i)\rho(\tau)U_{\text{LZ}}(\tau,\tau_i),\\
{\tilde{\sigma}}^z(\tau,\tau_i)=&U^{\dagger}_{\text{LZ}}(\tau,\tau_i){\sigma}^zU_{\text{LZ}}(\tau,\tau_i),
\end{align}
the transition probability $\ev{\overline{P(\tau)}}$ and the time-evolution equation for $\overline{\tilde{\rho}(\tau)}$ are given by
\begin{align}
\ev{\overline{P(\tau)}}=\Tr \qty[U_{\text{LZ}}(\tau,\tau_i)\overline{\tilde{\rho}(\tau)}U_{\text{LZ}}^{\dagger}(\tau,\tau_i)P],
\end{align}
\begin{align}
\frac{d}{d\tau}\overline{\tilde{\rho}(\tau)}
=\frac{1}{2}\lambda\sqrt{\kappa}\qty[{\tilde{\sigma}}^z(\tau,\tau_i),\qty[\overline{\tilde{\rho}(\tau)},{\tilde{\sigma}}^z(\tau,\tau_i)]].
\label{eq:mastereq_int}
\end{align}
Notice that the evolution operator $U_{\text{LZ}}$ and ${\tilde \sigma}^z$ are noise-independent. By integrating Eq.~\eqref{eq:mastereq_int} over $\tau$, the transition probability is given up to the first order in $\lambda$ by
\begin{widetext}
\begin{align}
\ev{\overline{P(\tau)}}=&\bra{\uparrow}U_{\text{LZ}}(\tau,\tau_i)\overline{\tilde{\rho}(\tau)}U_{\text{LZ}}^{\dagger}(\tau,\tau_i)\ket{\uparrow} \\
\simeq&\bra{\uparrow}U_{\text{LZ}}(\tau,\tau_i)\rho(\tau_i)U_{\text{LZ}}^{\dagger}(\tau,\tau_i)\ket{\uparrow}+\frac{\lambda\sqrt{\kappa}}{2}
\int^{\tau}_{\tau_i}d\tau'\bra{\uparrow}U_{\text{LZ}}(\tau,\tau_i)[{\tilde{\sigma}}^z(\tau',\tau_i),[\rho(\tau_i),{\tilde{\sigma}}^z(\tau',\tau_i)]]U_{\text{LZ}}^{\dagger}(\tau,\tau_i)\ket{\uparrow}.
\label{eq:approx_1st_numerical}
\end{align}
Substituting the specific form of $U_{\text{LZ}}$ and taking the limits $\tau_i\to-\infty$ and $\tau\to\infty$, the expression can be represented as
\begin{align}\label{eq:approx_1st}
\ev{\overline{P(\infty)}}\simeq e^{-2\pi\kappa}+&4\lambda\sqrt{\kappa}\int^{\infty}_{-\infty}d\tau\ \qty(|X_{\kappa}(\tau)|^2+\frac{2e^{-2\pi\kappa}}{1-e^{-2\pi\kappa}}\qty(\operatorname{Re}(X_{\kappa}(\tau)))^2+\frac{2e^{-\pi\kappa}}{1-e^{-2\pi\kappa}}\operatorname{Re}\qty(X_{\kappa}(\tau)) Y_{\kappa}(\tau)),
\end{align}
\begin{align}
X_{\kappa}(\tau)=&\kappa e^{-\frac{\pi\kappa}{2}}D_{-i\kappa-1}(e^{\frac{\pi}{4}i}\tau)D_{i\kappa-1}(-e^{-\frac{\pi}{4}i}\tau), \\
Y_{\kappa}(\tau)=&\frac{\kappa}{2}e^{-\frac{\pi\kappa}{2}}\qty(|D_{-i\kappa-1}(e^{\frac{\pi}{4}i}\tau)|^2+|D_{i\kappa-1}(-e^{-\frac{\pi}{4}i}\tau)|^2).
\end{align}
\end{widetext}
Both $|X_{\kappa}(\tau)|$ and $|Y_{\kappa}(\tau)|$ do not exceed 1. The derivation of Eq.~\eqref{eq:approx_1st} is given in Appendix \ref{sec:append_1st}. We refer to the transition probability numerically computed with Eq.~\eqref{eq:mastereq_int} as ``Numerical'', and those calculated up to the first order of $\lambda$ with Eq.~\eqref{eq:approx_1st_numerical} as ``First-order''. These two numerical plots are illustrated in Fig.~\ref{figure/rho_time_compare}. According to this figure, the transition probabilities with the first-order approximation approach those of the numerical simulations when $\kappa$ is small. As $\kappa$ increases, the accuracy of the approximation deteriorates due to the failure of the condition $\lambda\kappa\ll 1$.

We focus on an approximate solution at the final time $\tau_f\to\infty$ and seek the approximate solution for Eq.~\eqref{eq:approx_1st} in the regions (i) and (ii). First, we begin with the region (i) : small $\kappa$. Since $\lambda$ is sufficiently small, the first term on the right-hand side of Eq.~\eqref{eq:approx_1st} is dominant. Appendix \ref{sec:dominant} discusses how small $\lambda$ needs to be for the first term to dominate. Consequently, we can approximate the expression as
\begin{align}
\ev{\overline{P(\infty)}}\stackrel{\ri}{\simeq} e^{-2\pi\kappa}.
\label{eq:approx(i)}
\end{align}
This is strongly suppressed where $\kappa\gg 1$, and it only has significant values in the small $\kappa$ region. 

Second, consider the region (ii) : intermediate $\kappa$. Given the condition $\kappa\gg 1$, dropping terms proportional to $e^{-\pi \kappa}$ is an effective approximation. Under this approximation, only the term with $\qty|X_{\kappa}|^2$ in Eq.~\eqref{eq:approx_1st} contributes. In Appendix \ref{sec:dominant}, numerical computations confirm the dominance of the term with $\qty|X_{\kappa}|^2$ over the other terms. Thus, the transition probability at the final time can be approximated as
\begin{align}
&\quad \ev{\overline{P(\infty)}}\\
&\simeq \ 4\lambda\sqrt{\kappa}e^{-\pi\kappa}\kappa^2\\
&\quad \times \int^{\infty}_{-\infty}d\tau\ \qty|D_{-i\kappa-1}(e^{\frac{i\pi}{4}}\tau)D_{-i\kappa-1}(-e^{\frac{i\pi}{4}}\tau)|^2.
\label{eq:b1}
\end{align}
Ignoring the term of $\mathcal{O}(\kappa^{-2})$ and $\mathcal{O}(e^{-\pi\kappa})$, the transition probability is approximated as
\begin{align}
\ev{\overline{P(\infty)}}\stackrel{\rii}{\simeq} 2\pi\lambda\kappa.
\label{eq:approx(ii)}
\end{align}
The detailed derivation is conducted in Appendixes \ref{sec:append_approx1} and \ref{sec:append_approx2}. Equation~\eqref{eq:approx(ii)} approaches zero when $\kappa$ becomes zero and only has significant values in the large $\kappa$ region.

Therefore, the effective solution in the small $\kappa$ region~\eqref{eq:approx(i)} is dominant only in the small region. On the other hand, the effective solution in the large $\kappa$ region~\eqref{eq:approx(ii)} is dominant only in this large region. Hence, the transition probability effective in both regions (i) and (ii) can be approximated as the sum of the equations \eqref{eq:approx(i)} and \eqref{eq:approx(ii)}, 
\begin{align}
\ev{\overline{P(\infty)}}\stackrel{\ri,\rii}{\simeq}e^{-2\pi\kappa}+2\pi\lambda\kappa.
\label{eq:approx(i)(ii)}
\end{align}
The first term on the right-hand side represents the transition probability of the Landau--Zener model in the absence of noise. The second term corresponds to the influence of the noise. Despite being a transition probability, the right-hand side diverges as $\kappa\to\infty$. This is because this approximate solution is valid only in the region where $\lambda\kappa\ll 1$ and it is not allowed to take the limit $\kappa\to \infty$. This is confirmed by the comparison between numerical and approximate solutions in Fig.~\ref{figure/rho_time_compare}. As $\kappa$ increases, the influence of higher-order contributions of $\lambda$ becomes significant, leading to a deterioration in the accuracy of the first-order approximation.

\subsection{Adiabatic approximation}\label{subsec:adiabatic}
In the regions (ii) and (iii), that is $\kappa\gg 1$, the adiabatic approximation is employed as in~\cite{gulacsi2021defect}. First, we summarize the general theory of the adiabatic approximation in the time-dependent master equation for the two-level density matrix. A linear equation for the density matrix $\overline{\rho(\tau)}$ like the master equation~\eqref{eq:mastereq} is expressed as
\begin{align}
\frac{d}{d\tau}\dket{\rho(\tau)}
=\mathcal{L}(\tau)\dket{\rho}, \label{eq:master}
\end{align}
\begin{align}
\dket{\rho(\tau)}=\qty(\rho_{11}(\tau)\ \rho_{12}(\tau)\ \rho_{21}(\tau)\ \rho_{22}(\tau))^T,
\end{align}
\begin{align}
\overline{\rho(\tau)}=
\begin{pmatrix}
\rho_{11}(\tau) & \rho_{12}(\tau) \\
\rho_{21}(\tau) & \rho_{22}(\tau)
\end{pmatrix}.
\end{align}
Here, $\mathcal{L}(\tau)$ denotes the Liouvillian at time $\tau$, which is non-Hermitian $4\times4$ matrix. Diagonalizing $\mathcal{L}$ with a $4\times 4$ diagonalization matrix $S$ yields
\begin{align}
\operatorname{diag}(\chi_1(\tau),\chi_2(\tau),\chi_3(\tau),\chi_4(\tau))=S^{-1}(\tau)\mathcal{L}(\tau)S(\tau).
\end{align}
Let $\dket{\chi_\alpha(\tau)}$ and $\dbra{\hat{\chi}_\alpha(\tau)}$ be normalized right eigenvectors and left eigenvectors of $\mathcal{L}(\tau)$, satisfying
\begin{align}
\mathcal{L}(\tau)\dket{\chi_\alpha(\tau)}&=\chi_\alpha(\tau) \dket{\chi_\alpha(\tau)},\\
\dbra{\hat{\chi}_\alpha(\tau)}\mathcal{L}(\tau)&=\dbra{\hat{\chi}_\alpha(\tau)}\chi_\alpha(\tau),
\end{align}
with $\alpha=1,2,3,4$. When there is no degeneracy in the eigenvalues of the Liouvillian, i.e. $\chi_\alpha\neq \chi_\beta$, the condition $\dbraket{\hat{\chi}_\alpha(\tau)}{\chi_\beta(\tau)}=0$ is satisfied. The eigenvectors are normalized so that $\dbraket{\hat{\chi}_\alpha(\tau)}{\chi_\beta(\tau)}=\delta_{\alpha\beta}$. Then, the matrix $S$ and the inverse matrix $S^{-1}$ can be obtained as
\begin{align}
S(\tau)=\qty(\dket{\chi_1(\tau)}\ \dket{\chi_2(\tau)}\ \dket{\chi_3(\tau)}\ \dket{\chi_4(\tau)}),
\end{align}
\begin{align}
S^{-1}(\tau)=
\begin{pmatrix}
\dbra{\hat{\chi}_1(\tau)}\\ \dbra{\hat{\chi}_2(\tau)}\\ \dbra{\hat{\chi}_3(\tau)}\\
\dbra{\hat{\chi}_4(\tau)}
\end{pmatrix}.
\end{align}
Expanding the density matrix in terms of the right eigenvectors as
\begin{align}
\dket{\rho(\tau)}=\sum_{\beta=1}^4 c_{\beta}(\tau) \dket{\chi_\beta(\tau)},
\label{eq:expansion}
\end{align}
the master equation~\eqref{eq:master} leads to
\begin{align}
\frac{d}{d\tau}{c}_{\alpha}(\tau)=\chi_\alpha(\tau)c_{\alpha}(\tau)-\sum_{\beta=1}^4 c_{\beta}(\tau)\dbra{\hat{\chi}_\alpha(\tau)}\frac{d}{d\tau}\dket{\chi_\beta(\tau)}.
\end{align}
Therefore, if $\dbra{\tilde{\chi}_\alpha(\tau)}\frac{d}{d\tau}\dket{\chi_\beta(\tau)}$ is sufficiently small, the equation becomes closed for $\alpha$, resulting in adiabatic time evolution. Therefore, the adiabatic condition is given by~\cite{sarandy2005adiabatic}
\begin{align}
\label{eq:adiabatic_condition}
l_{\alpha\beta}\ll r_{\alpha\beta},
\end{align}
\begin{align}
l_{\alpha\beta}=&\max_{\tau}\qty|\dbra{\hat{\chi}_\alpha(\tau)}\frac{d}{d\tau}\dket{\chi_\beta(\tau)}|,\\
r_{\alpha\beta}=&\min_{\tau}\qty|\chi_{\alpha}(\tau)-\chi_{\beta}(\tau)|.
\end{align}
Here, $r_{\alpha\beta}=r_{\beta\alpha}$. From Eq.~\eqref{eq:expansion}, the adiabatic basis is defined as
\begin{align}
\dket{\tilde{\rho}(\tau)}=S^{-1}(\tau)\dket{\rho(\tau)}=
\begin{pmatrix}
c_{1}(\tau)\
c_{2}(\tau)\
c_{3}(\tau)\
c_{4}(\tau)
\end{pmatrix}^T,
\end{align}
Under the adiabatic approximation, if we further impose the condition $\dbra{\hat{\chi}_\alpha(\tau)}\frac{d}{d\tau}\dket{\chi_\alpha(\tau)}=0$ for any $\alpha$, the master equation is expressed as
\begin{align}
\frac{d}{d\tau}{c}_{\alpha}(\tau)\simeq \chi_\alpha(\tau)c_{\alpha}(\tau).
\end{align}
Therefore, $c_{\alpha}$ is expressed as
\begin{align}
c_{\alpha}(\tau)\simeq c_{\alpha}(\tau_i) \exp\qty(\int^{\tau}_{\tau_i}d\tau' \chi_{\alpha}(\tau')).
\label{eq:integration_tau}
\end{align}
Under this approximation, by calculating $c_{\alpha}(\tau)$ and $\dket{\rho(\tau)}=S(\tau)\dket{\tilde{\rho}(\tau)}$, the approximate solution for $\dket{\rho(\tau)}$ can be obtained.

Next, we apply the adiabatic approximation to the system in this study and calculate the transition probability $\rho_{11}(\infty)$. By letting 
$z(\tau) = \frac{\tau}{2\sqrt{\kappa}}$, the Liouvillian from Eq.~\eqref{eq:mastereq} is expressed as
\begin{align}
\mathcal{L}(\tau)=&
\begin{pmatrix}
0&i\sqrt{\kappa}&-i\sqrt{\kappa}&0\\
i\sqrt{\kappa}&-i\tau-2\lambda\sqrt{\kappa}&0&-i\sqrt{\kappa}\\
-i\sqrt{\kappa}&0&i\tau-2\lambda\sqrt{\kappa}&i\sqrt{\kappa}\\
0&-i\sqrt{\kappa}&i\sqrt{\kappa}&0
\end{pmatrix}\\
=&\sqrt{\kappa}
\begin{pmatrix}
0&i&-i&0\\
i&-2iz-2\lambda&0&-i\\
-i&0&2iz-2\lambda&i\\
0&-i&i&0
\end{pmatrix}.
\end{align}
In the regime $\kappa\gg 1$, the time evolution with this Liouvillian is calculated with an adiabatic approximation. The eigenvalues of this Liouvillian are
\begin{align}
\chi_1=&0,\\
\chi_2=&-\sqrt{\kappa}\qty(\frac{2\lambda}{z^2+1}+\mathcal{O}(\lambda^3)),\\
\chi_3=&-\sqrt{\kappa}\qty(2i\sqrt{z^2+1}+\frac{2z^2+1}{z^2+1}\lambda+\mathcal{O}(\lambda^2)),\\
\chi_4=&-\sqrt{\kappa}\qty(-2i\sqrt{z^2+1}+\frac{2z^2+1}{z^2+1}\lambda+\mathcal{O}(\lambda^2)).
\end{align}
The term $\mathcal{O}(\lambda^2),\ \mathcal{O}(\lambda^2)$ depends on $z$ but does not depend on $\kappa$. This term shall be neglected in the following analysis in this subsection. The corresponding right eigenvectors can be determined using perturbation theory in $\lambda$ as
\begin{align}
\dket{\chi_1(\tau)}&=
\begin{pmatrix}
\frac{1}{\sqrt{2}}\\
0\\
0\\
\frac{1}{\sqrt{2}}
\end{pmatrix},\\
\dket{\chi_2(\tau)}&\simeq
\begin{pmatrix}
\frac{z}{\sqrt{2(z^2+1)}}\\
\frac{1}{\sqrt{2(z^2+1)}}+i\lambda \frac{z\sqrt{z^2+1} }{\sqrt{2}(z^2+1)^2}\\
\frac{1}{\sqrt{2(z^2+1)}}-i\lambda \frac{z\sqrt{z^2+1} }{\sqrt{2}(z^2+1)^2}\\
-\frac{z}{\sqrt{2(z^2+1)}}
\end{pmatrix},\\
\dket{\chi_3(\tau)}&\simeq
\begin{pmatrix}
\frac{1}{2\sqrt{z^2+1}}+i\lambda\frac{4z^2+1}{8(z^2+1)^2}\\
-\frac{1}{2}\qty(1+\frac{z}{\sqrt{z^2+1}})+i\lambda\frac{3z+\sqrt{z^2+1}}{8(z^2+1)^2}\\
\frac{1}{2}\qty(1-\frac{z}{\sqrt{z^2+1}})+i\lambda\frac{3z-\sqrt{z^2+1}}{8(z^2+1)^2}\\
-\frac{1}{2\sqrt{z^2+1}}-i\lambda\frac{4z^2+1}{8(z^2+1)^2}
\end{pmatrix},\\
\dket{\chi_4(\tau)}&\simeq
\begin{pmatrix}
\frac{1}{2\sqrt{z^2+1}}-i\lambda\frac{4z^2+1}{8(z^2+1)^2}\\
\frac{1}{2}\qty(1-\frac{z}{\sqrt{z^2+1}})-i\lambda\frac{3z-\sqrt{z^2+1}}{8(z^2+1)^2}\\
-\frac{1}{2}\qty(1+\frac{z}{\sqrt{z^2+1}})-i\lambda\frac{3z+\sqrt{z^2+1}}{8(z^2+1)^2}\\
-\frac{1}{2\sqrt{z^2+1}}+i\lambda\frac{4z^2+1}{8(z^2+1)^2}
\end{pmatrix}.
\end{align}
Here, these eigenvectors satisfy $l_{\alpha\beta}=l_{\beta\alpha}$ and $l_{\alpha\alpha}=0$. The right eigenvector $\dket{\chi_1(\tau)}$ is considered to be the infinite-temperature state, which represents the state that both the probability of transitioning and the probability of not transitioning are ${1}/{2}$.

Now, we verify the adiabatic condition~\eqref{eq:adiabatic_condition}. The matrix representation of $\dbra{\hat{\chi}_\alpha(\tau)}\frac{d}{d\tau}\dket{\chi_\beta(\tau)}$ in $\alpha$ row and $\beta$ column is expressed as
\begin{widetext}
\begin{align}
\dbra{\hat{\chi}_\alpha(\tau)}\frac{d}{d\tau}\dket{\chi_\beta(\tau)}
\simeq\frac{1}{2\sqrt{\kappa}}\begin{pmatrix}
0&0&0&0\\
0&0&-\frac{1}{\sqrt{2}(z^2+1)}-i\lambda\frac{(8z^2-3)}{4\sqrt{2}(z^2+1)^{\frac{5}{2}}}&-\frac{1}{\sqrt{2}(z^2+1)}+i\lambda\frac{(8z^2-3)}{4\sqrt{2}(z^2+1)^{\frac{5}{2}}}\\
0&\frac{1}{\sqrt{2}(z^2+1)}+i\lambda\frac{(8z^2-3)}{4\sqrt{2}(z^2+1)^{\frac{5}{2}}}&0&-i\lambda\frac{z}{4(z^2+1)^{\frac{5}{2}}}\\
0&\frac{1}{\sqrt{2}(z^2+1)}-i\lambda\frac{(8z^2-3)}{4\sqrt{2}(z^2+1)^{\frac{5}{2}}}&i\lambda\frac{z}{4(z^2+1)^{\frac{5}{2}}}&0
\end{pmatrix}.
\end{align}
\end{widetext}
From $l_{12}=l_{13}=l_{14}=0$, the state $\dket{\chi_{1}}$ is adiabatic. In other states, the following results are obtained. $l_{23},\ l_{24}\simeq\frac{1}{2\sqrt{2\kappa}}$ and $r_{23},\ r_{24}\simeq2\sqrt{\kappa}$. $l_{34}\simeq\frac{2\lambda}{25\sqrt{5\kappa}}$ and $r_{34}\simeq4\sqrt{\kappa}$. Therefore, the adiabatic condition is $\kappa\gg 1$.

When the adiabatic condition is satisfied, we determine the transition probability. The initial condition in the original basis is $\dket{\rho(\tau_i)}=(1\ 0\ 0\ 0)^T$. In the limit $\tau_i\to-\infty$, the initial conditions in the adiabatic basis are given by $c_1(\tau_i)\to\frac{1}{\sqrt{2}},\ c_2(\tau_i)\to-\frac{1}{\sqrt{2}},\ c_3(\tau_i)\to0,\ c_4(\tau_i)\to0$ in our choice of eigenvectors. Then, the density matrix $\dket{\rho(\tau)}=S(\tau)\dket{\tilde{\rho}(\tau)}$ can be approximated as
\begin{widetext}
\begin{align}
\dket{\rho(\tau)}\simeq S(\tau)
\begin{pmatrix}
\frac{1}{\sqrt{2}}\\-\frac{1}{\sqrt{2}}\exp\qty(-\int^\tau_{\tau_i}d\tau' \chi_2\qty(\tau')) \\0 \\0
\end{pmatrix}
=\frac{1}{2}
\begin{pmatrix}
1-\frac{z(\tau)}{\sqrt{z^2(\tau)+1}}\exp\qty(-\int^\tau_{\tau_i}d\tau' \frac{2\lambda\sqrt{\kappa}}{z^2(\tau')+1}) \\
-\qty(\frac{1}{\sqrt{z^2(\tau)+1}}+i\lambda\frac{z(\tau)}{(z^2(\tau)+1)^{\frac{3}{2}}})\exp\qty(-\int^\tau_{\tau_i}d\tau' \frac{2\lambda\sqrt{\kappa}}{z^2(\tau')+1}) \\
-\qty(\frac{1}{\sqrt{z^2(\tau)+1}}-i\lambda\frac{z(\tau)}{(z^2(\tau)+1)^{\frac{3}{2}}})\exp\qty(-\int^\tau_{\tau_i}d\tau' \frac{2\lambda\sqrt{\kappa}}{z^2(\tau')+1})\\
1+\frac{z(\tau)}{\sqrt{z^2(\tau)+1}}\exp\qty(-\int^\tau_{\tau_i}d\tau' \frac{2\lambda\sqrt{\kappa}}{z^2(\tau')+1}).
\end{pmatrix}.
\end{align}
\end{widetext}
The density matrix at the final time is determined by substituting $\tau_i\to -\infty$, $\tau_f \to \infty$, resulting in
\begin{align}
\dket{\rho(\infty)}\simeq
\begin{pmatrix}
\frac{1}{2}\qty(1- \exp\qty(-4\pi\lambda\kappa))\\
0\\
0\\
\frac{1}{2}\qty(1+ \exp\qty(-4\pi\lambda\kappa))
\end{pmatrix}.
\end{align}
Here, the integral of $\chi_2(\tau)$ within the exponent is performed as
\begin{align}
\int^{\infty}_{-\infty}\frac{d\tau'}{z^2\qty(\tau')+1}=2\sqrt{\kappa}\int^{\infty}_{-\infty}\frac{dz}{z^2+1}=2\pi\sqrt{\kappa}.
\label{eq:integration_Z}
\end{align}
Here, a factor of $2\sqrt{\kappa}$ is introduced due to the Jacobian resulting from changing the integration variable from $d\tau$ to $dz$. Under the adiabatic approximation, the transition probability $\ev{\overline{P(\infty)}}=\rho_{11}(\infty)$ is given by
\begin{align}
\ev{\overline{P(\infty)}}\stackrel{\rii,\riii}{\simeq} \frac{1}{2}\qty(1- \exp\qty(-4\pi\lambda\kappa)).
\label{eq:approx(ii)(iii)}
\end{align}
In the derivation above, we have assumed that $\lambda$ is small and neglected the $\mathcal{O}(\lambda^3)$ terms within the exponent. However, when $\kappa$ is large, the approximate solution~\eqref{eq:approx(ii)(iii)} remains valid to all orders of $\lambda$. The reason for this is explained in Appendix \ref{sec:append_adiabatic}.

\subsection{Approximate solution for the transition probability}\label{subsec:all}
\begin{figure}[H]
\centering
\includegraphics[width=85mm]{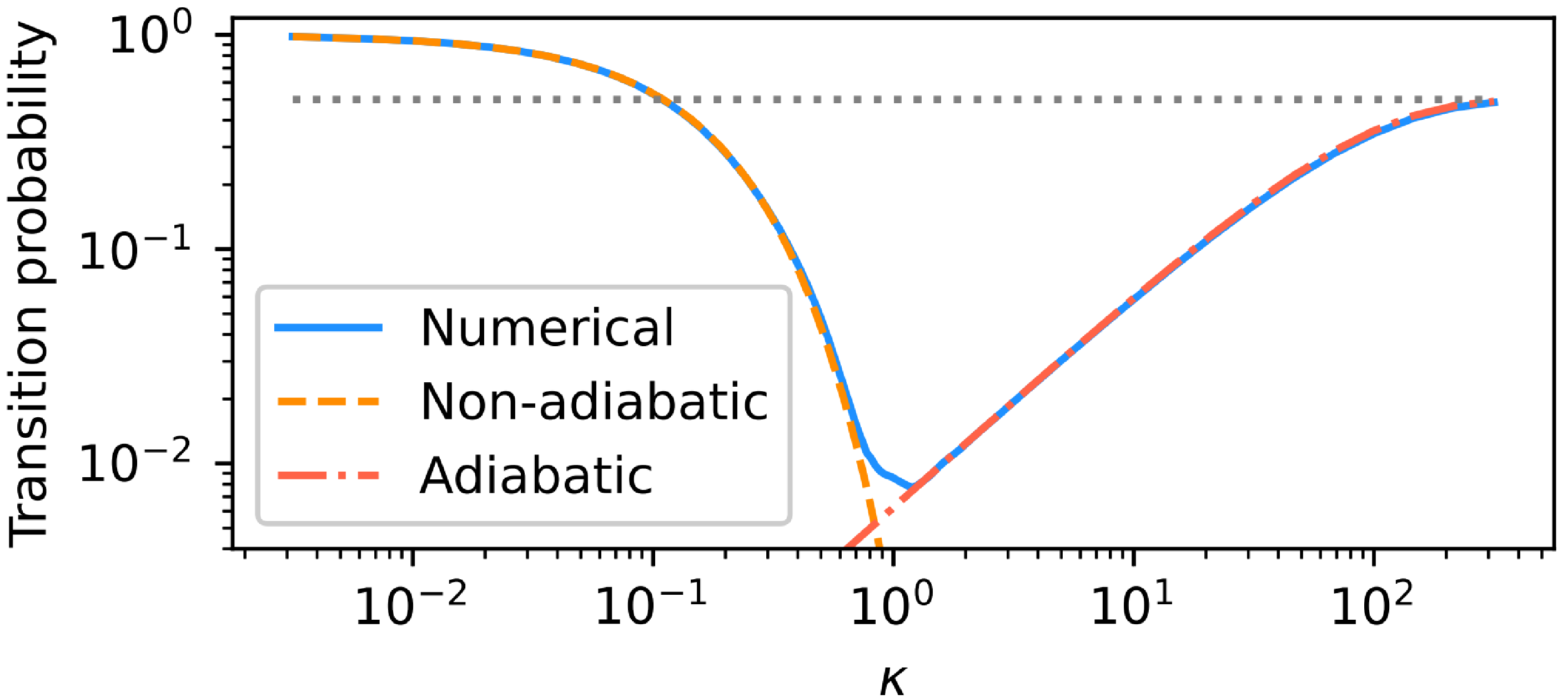}
\caption{Plots of the transition probabilities at $\tau_f$ against $\kappa$. The blue line (Numerical) represents the numerical simulation of Eq.~\eqref{eq:mastereq_int} over the time span from $\tau_i=-200$ to $\tau_f=200$. The yellow dashed line (Non-adiabatic) represents $p^{\text{non-ad}}$ and the red dashed line (Adiabatic) represents $p^{\text{ad}}$, with $\lambda=10^{-3}$. The gray dot line represents $1/2$.}
\label{figure/probability_kappa}
\end{figure}

Combining Eq.~\eqref{eq:approx(i)(ii)} and \eqref{eq:approx(ii)(iii)}, the effective transition probability in the regions (i), (ii), and (iii) is expressed as
\begin{align}
\label{eq:approx(i)(ii)(iii)}
\ev{\overline{P(\infty)}}\simeq p^{\text{non-ad}}(\kappa)+p^{\text{ad}}(\lambda,\kappa),
\end{align}
\begin{align}
p^{\text{non-ad}}(\kappa)
=e^{-2\pi\kappa},
\end{align}
\begin{align}
p^{\text{ad}}(\lambda,\kappa)=\frac{1}{2}\qty(1-\exp\qty(-4\pi\lambda\kappa)).
\end{align}
The first-order perturbation in $\lambda$ of Eq.~\eqref{eq:approx(i)(ii)(iii)} matches Eq.~\eqref{eq:approx(i)(ii)}. Figure~\ref{figure/probability_kappa} displays the numerical result obtained from solving the master equation~\eqref{eq:mastereq_int} and the approximate solution~\eqref{eq:approx(i)(ii)(iii)}. 

\begin{figure}[H]
\centering
\includegraphics[width=85mm]{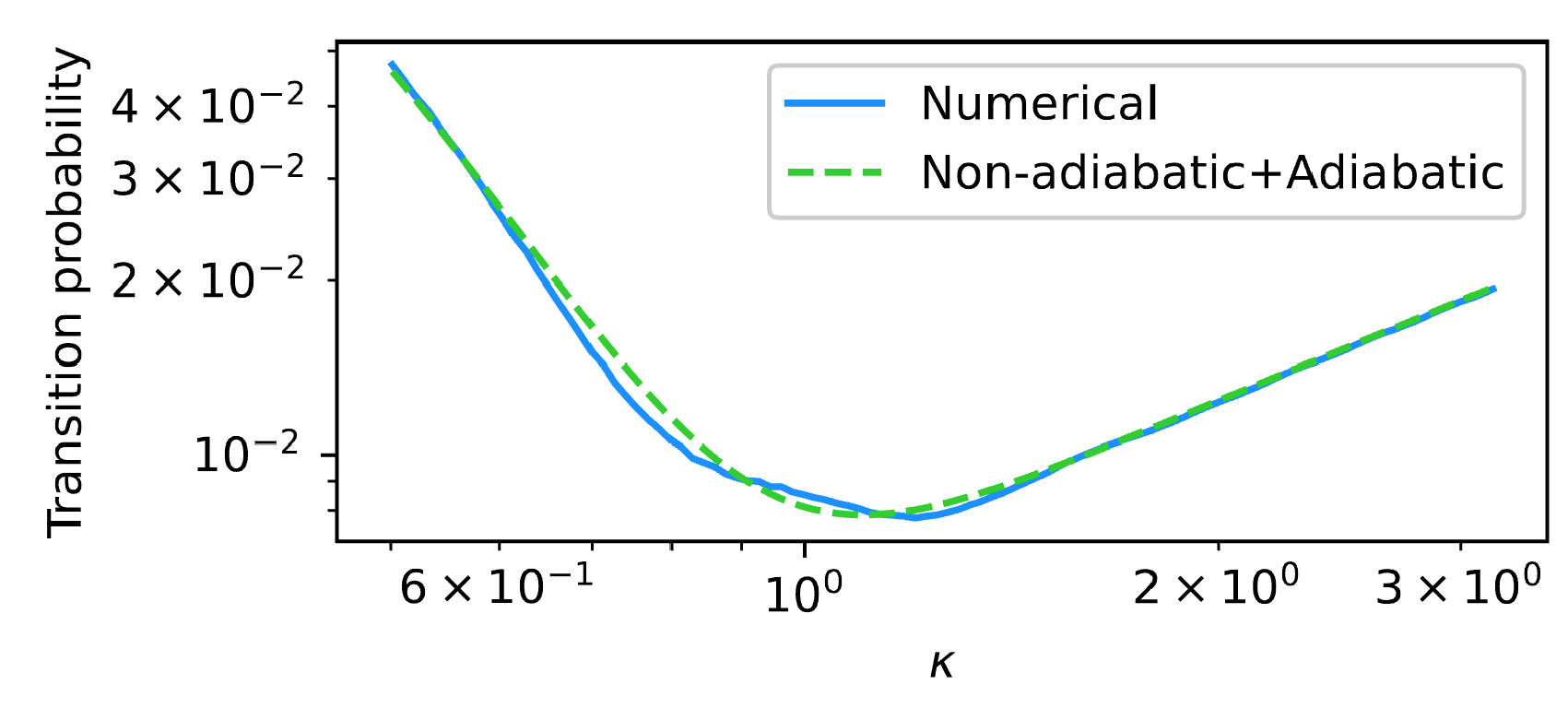}
\caption{Plots of the transition probabilities at $\tau_f$ against $\kappa$, focusing around $\kappa\approx1$. The blue line (Numerical) represents the numerical simulation of Eq.~\eqref{eq:mastereq_int} over the time interval from $\tau_i=-200$ to $\tau_f=200$. The green dashed line (Non-adiabatic+Adiabatic) represents $p^{\text{non-ad}}+p^{\text{ad}}$, with $\lambda=10^{-3}$.}
\label{figure/probability_kappa2}
\end{figure}

It is apparent that the non-adiabatic approximation is effective in the small $\kappa$ region, while the adiabatic approximate solution is effective in the large $\kappa$ region. We illustrate the behavior of the numerical result and the effective one $p^{\text{non-ad}}+p^{\text{ad}}$ in the $\kappa\simeq1$ domain in Fig.~\ref{figure/probability_kappa2}. 
While we have not derived a precise solution near $\kappa\simeq1$, it is confirmed that Eq.~\eqref{eq:approx(i)(ii)(iii)} quite well agree with the numerical result.

Consequently, it is confirmed that the transition probabilities in the noise-induced Landau-Zener model are expressed by Eq.~\eqref{eq:approx(i)(ii)(iii)}. This result corresponds to adding the adiabatic approximation result 
$p^{\text{ad}}$ to the well-known Landau-Zener transition probability $p^{\text{non-ad}}$.

The comparison with the study~\cite{kayanuma1984nonadiabatic} in the region where $\kappa$ is sufficiently large is discussed in Appendix \ref{sec:append_Kayanuma}.

\section{TRANSVERSE FIELD ISING model with Gaussian white noise}\label{sec:ising}
In Sec.~\ref{sec:SETUP}, We have considered the two-level Landau--Zener model with Gaussian white noise to analyze the transition probability. In this section, we apply this theoretical framework to the transverse field Ising model following the approach in~\cite{dziarmaga2005dynamics}. We aim to find an approximate solution for the defect density in the model with noise. The Hamiltonian is described as
\begin{align}
&{H}_{\text{Ising+noise}}(t)\\
=&{H}_{\text{Ising}}(t)+{H}_{\text{noise}}(t)\\
=&-\frac{1}{2}\sum^{N}_{j=1}\qty[\frac{vt}{2}{\sigma}^z_{j}+J{\sigma}^x_{j}{\sigma}^x_{j+1}]
-\frac{1}{2}\sum^{N}_{j=1}\gamma(t){\sigma}^z_{j}.
\label{eq:Hamiltonian}
\end{align}
Here, $\gamma(t)$ is Gaussian white noise that satisfies Eq.~\eqref{eq:white_noise}. If $\gamma(t)$ is absent, the system represents the transverse field Ising model, which is a model of a spin chain, where spins interact with the nearest-neighbor spins with $J$. The contribution of a uniform control field is expressed as $vt/2$. Here, ${\sigma}^a_j$ denote Pauli matrices acting on the $j$-th spin, with a periodic boundary condition ${\sigma}^a_{N+1}={\sigma}^a_{1}$. The number of spins $N$ is sufficiently large, corresponding to a thermodynamic limit. For simplicity, we assume $N$ is an even number. We focus on the adiabatic regime, where $\kappa=J^2/v$ is large. The initial state is set to be the ground state of the Hamiltonian ${H}_{\text{Ising}}$ and the state evolves under this Hamiltonian. The initial state at the initial time $t_i\to -\infty$ consists of spin-down states $\ket{\psi(t_i)}=\bigotimes_{j=1}^N \ket{\downarrow}_j$, where ${\sigma}^z_j\ket{\downarrow}_j=-\ket{\downarrow}_j$. The ground state at the final time $t_f\to \infty$ consists of spin-up states $\bigotimes_{j=1}^N \ket{\uparrow}_j$. Therefore, the defect density corresponds to the density of spin-down states at the final time. We introduce the spinless fermion operator ${c}_j$ and perform the Jordan-Wigner transformation~\cite{jordan1993paulische} as
\begin{align}
{\sigma}^z_j=1-2{c}^{\dagger}_j {c}_j,\quad {\sigma}^x_j=\qty({c}^{\dagger}_j+{c}_j)\prod_{l<j}\qty(-{\sigma}^z_l),
\end{align}
where $1\leq j\leq N$. The operators ${c}_j$ satisfy the anticommutation relations $\{{c}_j,{c}_k^{\dagger}\}=\delta_{jk}$ and $\{{c}_j,{c}_k\}=\{{c}_j^{\dagger},{c}_k^{\dagger}\}=0$. When $\prod_{j=1}^N{\sigma}^z_j$ acts on the initial state, the eigenvalue is 1. It indicates that ${c}_j$ follows the anti-periodic boundary condition ${c}_{N+1}=-{c}_1$~\cite{dziarmaga2005dynamics}. Therefore, performing the Fourier transform of the operator ${c}_j$, we use mode $q$ operator $\hat{c}_q$ as
\begin{align}
{c}_j=\frac{1}{\sqrt{N}} e^{-i\frac{\pi}{4}} \sum_q e^{-iqj}\hat{c}_q,
\end{align}
where $q=\pm (2 n-1) \pi/N, \ n\in{1, \cdots,N/2}$. $\hat{c}_q$ also satisfies the fermionic anticommutation relations. Then, the Hamiltonian can be expressed with the operators $\hat{c}_q$ as
\begin{align}
&{H}_{\text{Ising+noise}}(t)\\
=&\sum_{q>0} (\hat{c}^{\dagger}_{q}\ \hat{c}_{-q})
\qty({h}_{q}(t)+{h}_{\text{noise}}(t))
\mqty(\hat{c}_{q}\\\hat{c}^{\dagger}_{-q}),
\label{eq:Hamiltoniam_cq}
\end{align}
where
\begin{align}
h_q(t)= \qty(\frac{vt}{2}+J\cos q){\sigma}^z+{J}\sin q{\sigma}^x,
\end{align}
\begin{align}
\quad {h}_{\text{noise}}(t)=\gamma(t){\sigma}^z.
\end{align}
For $q>0$, we define the states $\ket{\uparrow}_q=\hat{c}_{q}^{\dagger}\hat{c}_{-q}^{\dagger}\ket{0}_q$ and $\ket{\downarrow}_q=\ket{0}_q$, where $\ket{0}_q$ represents the normalized vacuum state for mode $\pm q$, satisfying $\hat{c}_q\ket{0}_q=\hat{c}_{-q}\ket{0}_q=0$. The ground state at the initial time $t_i$ is represented as $\ket{\psi(t_i)}=\ket{\uparrow}_{q}^{\otimes{q>0}}$. Since the Hamiltonian is expressed as a sum over each mode $q$ and the total sum over $q$ remains conserved, the density operator at any time $t$ can be decomposed as
\begin{align}
\hat{\rho}(t)=\bigotimes_{q>0}\hat{\rho}_{q}(t).
\end{align}
Here, $\hat{\rho}_{q}$ is a density operator constructed only from $\ket{\uparrow}_{q}$ and $\ket{\downarrow}_{q}$. For the initial conditions in mode $q$, the pure ground state is set as $\hat{\rho}_{q}\qty(t_i)=\ket{\uparrow}_q\bra{\uparrow}$. We represents the density operator $\hat{\rho}_{q}(t)$ in matrix form as
\begin{align}
\rho_{q}(t)=
\begin{pmatrix}
{}_q\bra{\uparrow}\hat{\rho}_{q}(t)\ket{\uparrow}_q&{}_q\bra{\uparrow}\hat{\rho}_{q}(t)\ket{\downarrow}_q\\
{}_q\bra{\downarrow}\hat{\rho}_{q}(t)\ket{\uparrow}_q&{}_q\bra{\downarrow}\hat{\rho}_{q}(t)\ket{\downarrow}_q
\end{pmatrix}.
\end{align}
The density matrix $\rho_q(t)$ evolves under the effective Hamiltonian ${h}_{q}(t)+{h}_{\text{noise}}(t)$. The transition probability in mode $q$ at time $t$ is expressed using the time evolution operator $U_q(t,t_i)$ and the noise-averaged density matrix $\overline{\rho_{q}(t)}$ as
\begin{align}
\ev{\overline{P_q(t)}}=&\Tr\qty[\overline{\rho_{q}(t)}P_q]\\
=&\Tr[\overline{U_q(t,t_i)\rho_q(t_i)U_q^{\dagger}(t,t_i)}P_q],
\end{align}
where $P_q=\text{diag}(1,0)$, which corresponds to $\ket{\uparrow}_q\bra{\uparrow}$. Similarly to Sec.~\ref{sec:SETUP}, the noise-averaged density matrix $\overline{\rho_q(t)}$ satisfies the master equation
\begin{align}
\frac{d}{dt}\overline{\rho_q(t)}=-i\qty[h_q(t),\overline{\rho_q(t)}]+\frac{1}{2}W^2\qty[\qty[{\sigma}^z,\overline{\rho_q(t)}],{\sigma}^z].
\end{align}
The differences from Eq.~\eqref{eq:LZ_noise} lies in a shift $J\cos q$, and the modification of the coupling term from $J$ in ${H}_{\text{LZ}}$ to $J\sin q$ in $h_q$. By considering the transition between remote past and future, any finite time shifts in $h_q$ can be neglected. Therefore, the dimensionless master equation is expressed as
\begin{align}
\frac{d}{d\tau}\overline{\rho_q(\tau)}
=&-i\qty[\tilde{h}_q(\tau),\overline{\rho_q(\tau)}]\\
&+\frac{\lambda}{2}\sqrt{\kappa}\qty[\qty[{\sigma}^z,\overline{\rho_q(\tau)}],{\sigma}^z],
\label{eq:mastereq2}
\end{align}
\begin{align}
\tilde{h}_q(\tau)&=\frac{\tau}{2}{\sigma}^z+\sqrt{\kappa_q}{\sigma}^x,\\
\kappa_q &=\kappa\sin^2 q,
\end{align}
where $\kappa=J^2/v$ is the adiabatic parameter and $\lambda=W^2/J$ is the noise strength parameter. In the following, we consider $\kappa\gg1$. This master equation~\eqref{eq:mastereq2} is similar to Eq.~\eqref{eq:mastereq}. The parameter $\kappa_q$ satisfies the conditions $\kappa_q \leq \kappa$ and $\kappa_{\frac{\pi}{2}}=\kappa$. In the vicinity of $q\approx 0,\ \pi$, the energy gap decreases, indicating the non-adiabatic regime. In contrast, in other regions, a significant energy gap remains, indicating the adiabatic domain. In the right-hand side of Eq.~\eqref{eq:mastereq2}, since $\lambda\sqrt{\kappa} = \lambda\sqrt{\kappa_q}/|\sin q|$, we can directly apply the approximate formula~\eqref{eq:approx(i)(ii)(iii)} by transforming $\kappa\to\kappa_q$ and $\lambda$ to $\lambda/|\sin q|$. For the numerical simulations, we consider the master equation in the interaction picture, corresponding to equation~\eqref{eq:mastereq_int}, given by
\begin{align}
\frac{d}{d\tau}\overline{\tilde{\rho}_q(\tau)}
=\frac{1}{2}\lambda\sqrt{\kappa}\qty[{\tilde{\sigma}}^z_q(\tau,\tau_i),\qty[\overline{\tilde{\rho}_q(\tau)},{\tilde{\sigma}}^z_q(\tau,\tau_i)]].
\label{eq:mastereq_int2}
\end{align}
The transformation to the interaction picture is defined as
\begin{align}
\tilde{\rho}_q(\tau)=&U^{\dagger}_{\text{LZ},q}(\tau,\tau_i)\rho_q(\tau)U_{\text{LZ},q}(\tau,\tau_i),\\
{\tilde{\sigma}}^z_q(\tau,\tau_i)=&U^{\dagger}_{\text{LZ},q}(\tau,\tau_i){\sigma}^zU_{\text{LZ},q}(\tau,\tau_i),
\end{align}
where $U_{\text{LZ},q}$ is obtained by replacing $\kappa$ in $U_{\text{LZ}}$ in Eq.~\eqref{eq:unitary} with $\kappa_q$.

The approximate solution of the transition probabilities of the mode $q$ is used to calculate the defect density. The defect number operator $\mathcal{N}$ is represented as
\begin{align}
\mathcal{N}=\frac{1}{2}\sum_j \qty(1-{\sigma}^z_j)=\sum_j {c}^{\dagger}_{j} {c}_{j}=\sum_q \hat{c}^{\dagger}_{q} \hat{c}_{q}.
\end{align}
Therefore, the number of defects is the sum of the particles present in q modes. As the state $\ket{\uparrow}_q$ is a two-particle state, the expected defects number is twice the transition probability. Consequently, the expected defect density at the final time can be expressed as
\begin{align}
\label{eq:defect}
n=\frac{\ev{\mathcal{N}(\tau_f)}}{N}=\frac{2}{N}\sum_{q>0}\ev{\overline{P_{q}(\tau_f)}}.
\end{align}
In the thermodynamic limit, the expectation value of the defect density can be approximated as
\begin{align}
n\to \int^{\pi}_{0}\frac{dq}{\pi} \ev{\overline{P_q(\tau_f)}},
\end{align}
where $\frac{1}{N}\sum_{q>0}\to\int^{\pi}_0\frac{dq}{2\pi}$ is used. By using the approximate solution for the Landau--Zener model~\eqref{eq:approx(i)(ii)(iii)}, and making the substitutions $\kappa \to \kappa_q$ and $\lambda\to\lambda/|\sin q|$, we obtain
\begin{align}
\label{eq:u^2}
\ev{\overline{P_q(\infty)}}&\simeq l^{\text{non-ad}}_{\kappa}(q)+l^{\text{ad}}_{\lambda,\kappa}(q),\\
l^{\text{non-ad}}_{\kappa}(q)&=e^{-2\pi\kappa\sin^2 q},\\
l^{\text{ad}}_{\lambda,\kappa}(q)&=\frac{1}{2}\qty(1-\exp\qty(-4\pi\lambda\kappa|\sin q|)).
\end{align}

\begin{figure}[H]
\centering
\includegraphics[width=85mm]{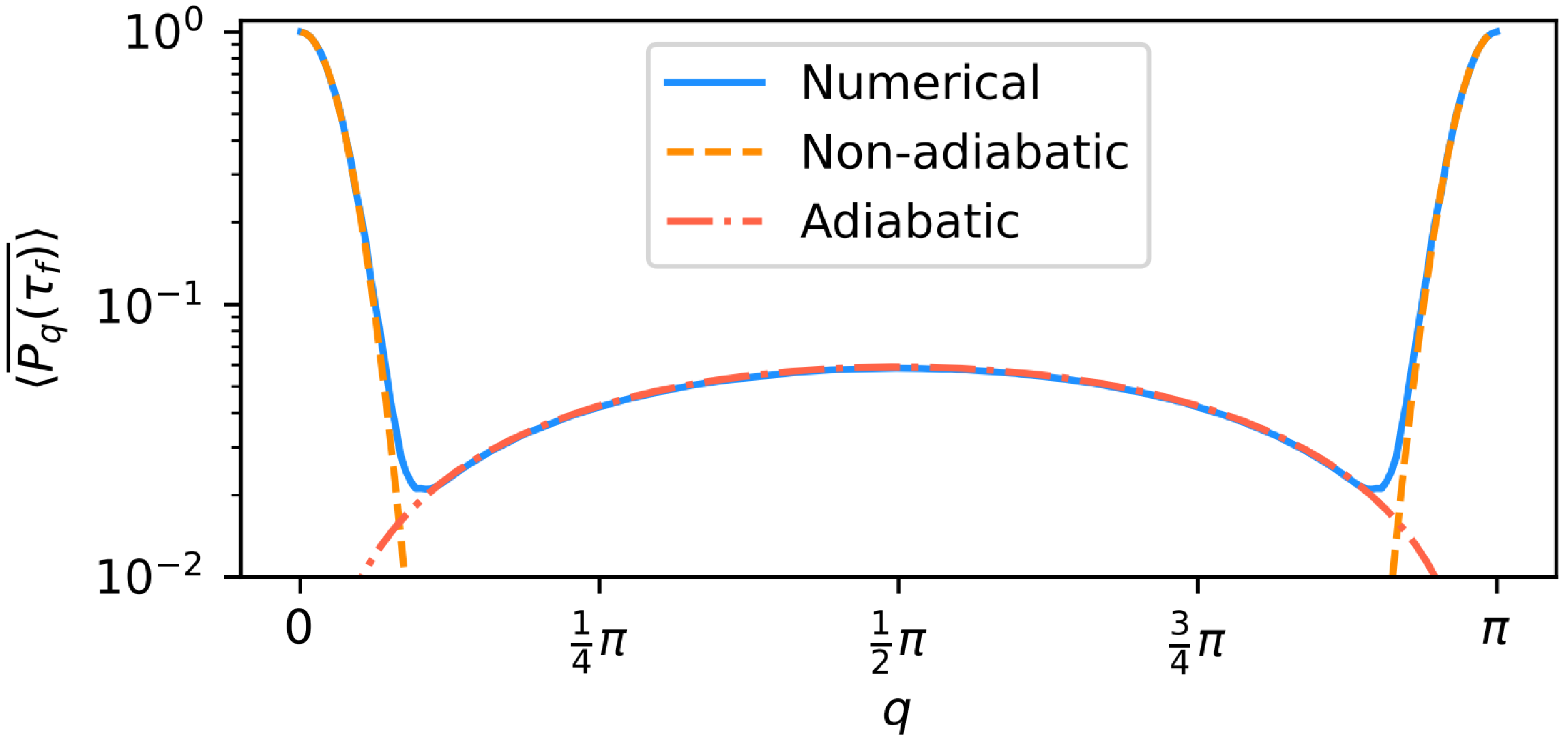}
\caption{Plots of the transition probability $\ev{\overline{P_q(\tau_f)}}$ against $q$. The blue line (Numerical) represents the numerical simulation of Eq.~\eqref{eq:mastereq_int2} over the time span from $\tau_i=-200$ to $\tau_f=200$. The yellow dashed line (Non-adiabatic) represents $l^{\text{non-ad}}_{\kappa}(q)$, while the red dashed line (Adiabatic) represents $l^{\text{ad}}_{\lambda,\kappa}(q)$, with $\kappa=10,\ \lambda=10^{-3}$.}
\label{figure/integrand1}
\end{figure}

We present the numerical results and the approximate solution~\eqref{eq:u^2} in Fig.~\ref{figure/integrand1}. The numerical result converges to $l^{\mathrm{non-ad}}_{\kappa}$ when $q$ is close to 0 or $\pi$, while it converges to $l^{ad}_{\lambda,\kappa}$ elsewhere. It shows that Eq.~\eqref{eq:u^2} provides a well-approximated solution and its integration yields an approximation of the defect.

Using Eq.~\eqref{eq:u^2}, the defect density $n(\lambda,\kappa)$ can be approximated as
\begin{align}
\label{eq:defect_infty}
n^{\infty\text{-order}}(\lambda,\kappa)=& n^{\text{KZM}}(\kappa)+n^{\text{noise}}(\lambda,\kappa),
\end{align}
\begin{align}
n^{\text{KZM}}(\kappa)
&=\int^{\pi}_{0}\frac{dq}{\pi} l^{\text{non-ad}}_{\kappa}(q)=\int^{\pi}_{0}\frac{dq}{\pi} e^{-2\pi\kappa\sin^2 q}\\
&=e^{-\pi\kappa}I_0(\pi\kappa)\simeq
\frac{1}{\pi\sqrt{2\kappa}},
\label{eq:n_KZM}
\end{align}
\begin{align}
n^{\text{noise}}(\lambda,\kappa)&=\int^{\pi}_{0}\frac{dq}{\pi} l^{\text{ad}}_{\lambda,\kappa}(q)\\
&=\int^{\pi}_{0}\frac{dq}{\pi} \qty(\frac{1}{2}-\frac{1}{2}\exp\qty(-4\pi\lambda\kappa|\sin q|))\\
&=\frac{1}{2}-\frac{1}{2}I_0\qty(4\pi \lambda\kappa)+\frac{1}{2}L_0\qty(4\pi \lambda\kappa)
\end{align}
Here, the condition $\kappa\gg1$ is used in the calculation in Eq.~\eqref{eq:n_KZM}. Consequently, the approximate solution $n^{\infty\text{-order}}$ is valid only in the regime $\kappa\gg1$. $n^{\infty\text{-order}}$ includes contributions all order in $\lambda$. $I_0(z)$ is the modified Bessel function of the first kind and $L_0(z)$ is the modified Struve function~\cite{gradshteyn2014table}. $n^{\text{KZM}}(\kappa)$ represents the defect density in the absence of noise and corresponds to the result of the KZM. $n^{\text{noise}}$ represents the contribution arising from Gaussian white noise. By using the asymptotic expansions of $I_0(z)$ and $L_0(z)$, $n^{\infty\text{-order}}\to1/2$ can be verified when $\lambda\kappa\to\infty$. This implies that the system asymptotically approaches the infinite-temperature state as the noise becomes large. In contrast, when $\lambda\kappa\ll1$, using the properties $I_0(z)=1+\frac{z^2}{4}+\mathcal{O}(z^4)$ and $L_0(z)=\frac{2z}{\pi}+\mathcal{O}(z^3)$, we can obtain a refined expression
\begin{align}
\label{eq:defect_1st}
n^{\infty\text{-order}}(\lambda,\kappa)\simeq \frac{1}{\pi \sqrt{2\kappa}}+4\lambda\kappa
=:n^{\text{1st}}(\lambda,\kappa).
\end{align}
The first term in Eq.~\eqref{eq:defect_1st} represents the defect density in the transverse field Ising model without noise. The second term in Eq.~\eqref{eq:defect_1st} represents the contribution of noise and is derived using the adiabatic approximation. This term is proportional to $v^{-1} \propto \kappa$, which is expected from computational researches~\cite{gao2017anti, dutta2016anti}. Then, Eq.~\eqref{eq:defect_1st} demonstrates the successful reproduction of the anti-KZM scaling previously known. Therefore, instead of tracking the time evolution of states, as done in previous research, the anti-KZM scaling can be directly obtained by applying the adiabatic approximation to the Liouvillian in the master equation.

\begin{figure}[H]
    \centering
\includegraphics[width=85mm]{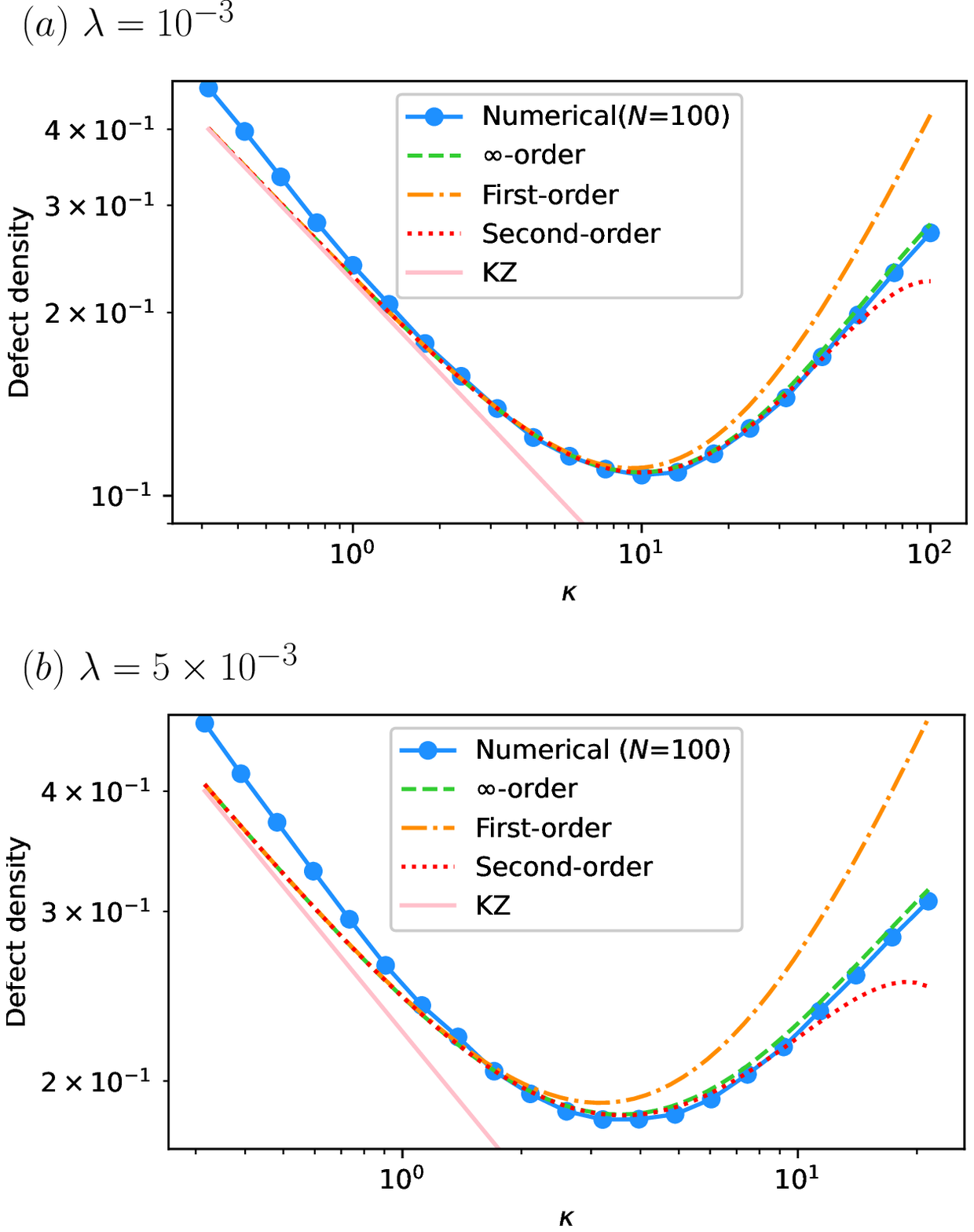}
\caption{Plots of the defect density against $\kappa$. The upper (a) corresponds to $\lambda=10^{-3}$, while the lower (b) corresponds to $\lambda=5\times10^{-3}$. The blue solid lines (Numerical) represent the numerical simulation of Eq.~\eqref{eq:mastereq_int2} over the time span $\tau_i=-200$ to $\tau_f=200$, plotting the discrete sum in Eq.~\eqref{eq:defect} with $N=100$. The green dashed lines ($\infty$-order) represent $n^{\infty\text{-order}}$, the yellow dashed lines (First-order) represent $n^{\text{1st}}$, and the red dot lines (Second-order) represent $n^{\text{2nd}}$. The pink solid lines (KZ) represent $n^{\text{KZM}}$.}
\label{figure/defect_scaling}
\end{figure}

The second term diverges as $\kappa\to \infty$, which is due to neglecting higher-order contributions of $\lambda\kappa$. Incorporating these higher-order $\lambda$ contributions improves the accuracy of the approximation. The approximate solution including contributions up to the second order of $\lambda$ is given by
\begin{align}
n^{\infty\text{-order}}(\lambda,\kappa)&\simeq\frac{1}{\pi \sqrt{2\kappa}}+4\lambda\kappa-2\pi^2\lambda^2\kappa^2\\
\label{eq:defect_2nd}
&=:n^{\text{2nd}}(\lambda,\kappa).
\end{align}
The defect density $n$ obtained numerically, along with $n^{\infty\text{-order}}$, $n^{\text{1st}}$, and $n^{\text{2nd}}$ are compared in Fig.~\ref{figure/defect_scaling}. In the numerical calculations, the number of spins $N$ is set to 100. $\lambda$ is smaller in the upper figure (a) than the lower figure (b).
It can be seen that $n^{\infty\text{-order}}$ closely matches the numerically obtained defect density in the region where $\kappa$ is large, indicating that it serves as a good approximation of the defect density in the adiabatic region. Hence, we have successfully obtained a new scaling $n^{\infty\text{-order}}$. Note that in the region where $\kappa$ is small, $n^{\infty\text{-order}}$ is not precise because the approximation~\eqref{eq:n_KZM} becomes less effective. Both $n^{\text{1st}}$ and $n^{\text{2nd}}$ serve as approximate solutions to $n^{\infty\text{-order}}$ when $\kappa$ is small. It is observed that the region where the $n^{\text{2nd}}$ is effective is larger than the region where $n^{\text{1st}}$ is effective. 

The parameter $\kappa_{\opt}$ is defined as the value of $\kappa$ that minimizes the defect density. As $\lambda$ increases, $\kappa_{\opt}$ decreases, which worsens the validity of $n^{\text{1st}}$ near $\kappa_{\opt}$. In this regime, $n^{\text{2nd}}$ becomes more effective than $n^{\text{1st}}$.

The behavior of the defect density in the regime where $\kappa$ is sufficiently large is provided in Appendix \ref{sec:append_Kayanuma}.

\section{Derivation of the optimization parameter}\label{sec:vopt}
\begin{figure}[H]
\centering
\includegraphics[width=85mm]{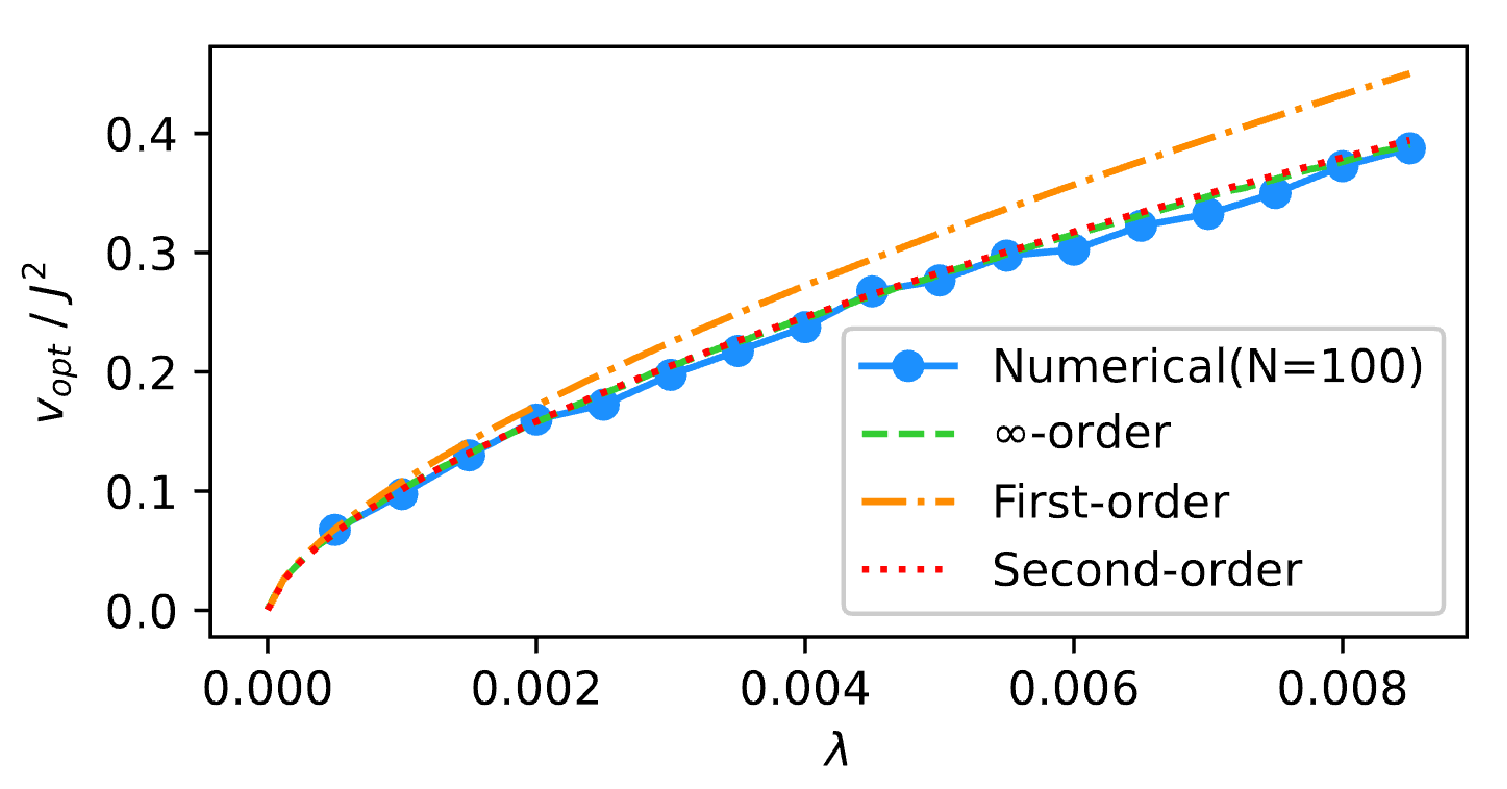}
\caption{Plots of $v_{\opt}$ against $\lambda$. The blue line (Numerical) represents the numerical simulation of Eq.~\eqref{eq:mastereq_int2} over the time span $\tau_i=-200$ to $\tau_f=200$, plotting the discrete sum in Eq.~\eqref{eq:defect}, which represents $v_{\opt}^{\num}$ with $N=100$. The green dashed line ($\infty$-order) represents $v^{\infty\text{-order}}_{\opt}$, the yellow dashed line (First-order) represents $v^{\text{1st}}_{\opt}$, and the red dot line (Second-order) represents $v^{\text{2nd}}_{\opt}$.}
\label{figure/v_opt}
\end{figure}

We shall find an optimization parameter that minimizes the defect density $n$. Let $v_{\opt}$ denote the driving speed $v$ at which the defect density attains its minimum. The previous studies~\cite{gao2017anti, dutta2016anti} numerically obtain the scaling of $v_{\opt}$ with respect to the parameter $W$, which represents the strength of noise. We seek the approximate solution of $v_{\opt}$ through the stationary point. The derivative of $n^{\text{1st}}$,
\begin{align}
\frac{d}{d\kappa}n^{\text{1st}}= -\frac{1}{\pi(2\kappa)^{\frac{3}{2}}}+4\lambda
\end{align}
vanishes at $\kappa=\kappa^{\text{1st}}_{\opt}=(2^7\pi^2\lambda^2)^{-\frac{1}{3}}$, which minimizes $n^{\text{1st}}$. Since $\kappa=J^2/v$, $v_{\opt}$ for $n^{\text{1st}}$ is given by
\begin{align}
v^{\text{1st}}_{\opt}/J^2=(2^7\pi^2\lambda^2)^{\frac{1}{3}}.
\end{align}
Similar to the previous study~\cite{dutta2016anti}, we have demonstrated $v_{\opt}^{\text{1st}}\propto \lambda^{2/3}\propto W^{4/3}$. Next, we determine $v^{\text{2nd}}$, which minimizes $n^{\text{2nd}}$. This is because comparing the results up to the 
$\lambda^2$ term allows us to quantitatively assess how much the results up to the $\lambda^1$ term deviate from the true results. Since the stationary point $\kappa^{\text{2nd}}$ cannot be determined analytically, we make an assumption that $\kappa^{\text{2nd}}$ is close to $\kappa^{\text{1st}}$ and determine $\kappa^{\text{2nd}}$ as the perturbation from $\kappa^{\text{1st}}_{\opt}$. Setting $\kappa^{\text{2nd}}_{\opt}=\kappa^{\text{1st}}_{\opt}\qty(1+\varepsilon)$, the derivative is expressed as
\begin{align}
&\left.\frac{d}{d\kappa}n^{\text{2nd}}\right|_{\kappa=\kappa^{\text{2nd}}_{\opt}}\\
=&-\frac{1}{\pi(2\kappa^{\text{2nd}}_{\opt})^{\frac{3}{2}}}+4\lambda-4\pi^2\lambda^2\kappa^{\text{2nd}}_{\opt}\\
=&-\frac{1}{\pi(2\kappa^{\text{1st}}_{\opt})^{\frac{3}{2}}}(1+\varepsilon)^{-\frac{3}{2}}+4\lambda-4\pi^2\lambda^2\kappa^{\text{1st}}_{\opt}(1+\varepsilon).
\end{align}
Given that $\varepsilon$ is sufficiently small, the perturbation up to the first order in $\varepsilon$ results in 
\begin{align}
\left.\frac{d}{d\kappa}n^{\text{2nd}}\right|_{\kappa=\kappa^{\text{2nd}}_{\opt}}\simeq\frac{3\varepsilon}{2\pi(2\kappa^{\text{1st}}_{\opt})^{\frac{3}{2}}}-4\pi^2\lambda^2\kappa^{\text{1st}}_{\opt}(1+\varepsilon).
\end{align}
For the right-hand side to be zero, $\varepsilon$ is given by
\begin{align}
\varepsilon=\frac{1}{3}\qty(\frac{\pi^4\lambda}{2^4})^{\frac{1}{3}}
+\mathcal{O}\qty(\lambda^{\frac{2}{3}}).
\end{align}
Thus, ignoring $\mathcal{O}\qty(\lambda^{4/3})$, $v_{\opt}$ for $n^{\text{2nd}}$ is approximated as
\begin{align}
v^{\text{2nd}}_{\opt}/J^2=(2^7\pi^2\lambda^2)^{\frac{1}{3}}
-\frac{2}{3}\pi^2\lambda.
\end{align}
By solving Eq.~\eqref{eq:mastereq_int2} for various values of $v$, the minimum value $v_{\opt}^{\num}$ is numerically obtained. If we set $n^{\infty\text{-order}}$ as an approximate solution for $n$, $v_{\opt}^{\infty\text{-order}}$ is numerically obtained by finding the minimum value as $v$ varies. $v_{\opt}^{\num}$, $v_{\opt}^{\infty\text{-order}}$, $v^{\text{1st}}_{\opt}$, and $v^{\text{2nd}}_{\opt}$ against $\lambda$ are plotted respectively in Fig.~\ref{figure/v_opt}.

It is confirmed that $v_{\opt}^{\infty\text{-order}}$ accurately reproduces $v_{\opt}$. Since $v^{\text{2nd}}_{\opt}$ approaches the numerical results more closely than $v^{\text{1st}}_{\opt}$. The relative deviation of $v^{\text{2nd}}_{\opt}$ over $v^{\text{1st}}_{\opt}$ is expressed as
\begin{align}
\frac{v^{\text{2nd}}_{\opt}}{v^{\text{1st}}_{\opt}} \simeq 1 - \left(\frac{\pi^4 \lambda}{2^4\times3^3}\right)^{\frac{1}{3}}=: 1 - \zeta(\lambda).
\end{align}
Let $n_{\opt}$ be the defect density when $v=v_{\opt}$, corresponding to the minimum value of $n$. The relative deviation is then expressed as
\begin{align}
\frac{n^{\text{2nd}}_{\opt}}{n^{\text{1st}}_{\opt}} \simeq 1 - \qty(\frac{\pi^4 \lambda}{2^{10}\times3^3})^{\frac{1}{3}}=: 1 - \xi(\lambda).
\end{align}
The magnitude of $\zeta$ and $\xi$ indicates the effectiveness of the approximation of $v^{\text{1st}}_{\opt}$ and $n^{\text{1st}}_{\opt}$ respectively. If both parameters are sufficiently small compared to 1, $v^{\text{1st}}_{\opt}$ and $n^{\text{1st}}_{\opt}$ can be considered effective. For example, when $\lambda=10^{-3}$, $\kappa^{\text{1st}}_{\opt}\approx9.3$, $\zeta\approx0.061$, and $\xi\approx0.015$. When $\lambda=5\times 10^{-3}$, $\kappa^{\text{1st}}_{\opt}\approx3.2$, $\zeta\approx0.10$, and $\xi\approx0.026$. We see that as $\lambda$ increases, $\zeta$ and $\xi$ increase. This leads to a loss of validity of the first-order approximation in the regime of large $\lambda$. Note that if $\lambda\gtrapprox 0.01$, the approximation formulas break down. This is because, as $\lambda$ increases, the stationary point of $\kappa$ decreases, making the approximation in Eq.~\eqref{eq:n_KZM} less effective near the stationary point.

\section{Conclusion}\label{sec:conclusion}
Under the assumption of small noise, $\lambda\ll1$, we analytically derived the transition probability of the Landau--Zener model with Gaussian white noise. We considered three $\kappa$ regions: (i) small, (ii) intermediate, and (iii) large. For the regions (i) and (ii), we solved a master equation under a first-order approximation for $\lambda$. For the regions (ii) and (iii), we employed an adiabatic approximation method.

Applying the analysis of the Landau--Zener model, we analytically derived the defect density in the transverse field Ising model with Gaussian white noise. The results reveal that in the regime where $\lambda\kappa\ll 1$, the anti-KZM scaling previously known through numerical computations~\cite{dutta2016anti, gao2017anti} is effective. On the other hand, outside the region where $\lambda\kappa\ll 1$, a new scaling was found. In previous derivations of the anti-KZM behavior of defect density, it has been necessary to solve differential equations. However, this results suggest that the anti-KZM behavior can be derived simply by applying the adiabatic approximation.

Furthermore, the optimal parameter $v_{\opt}$ follows the same scaling as previously known. It can be argued that to minimize the defect density, it is desirable to choose $v_{\opt}$ smaller than the value obtained from the anti-KZM previously known if the applied noise is large. We can evaluate the effectiveness of the first-order perturbation by comparing it with the second-order perturbation.

As an application to another system, we consider the Schwinger mechanism~\cite{sauter1931behavior,euler1936consequences,schwinger1951gauge}, where electron-positron pairs are expected to be generated when an electric field is applied to a vacuum. Since the time evolution of this system is related to the Landau--Zener system, we propose applying the methods of this research to derive the number density of electron-positron pairs generated by an electric field, under the influence of noise. It is anticipated that, due to the presence of noise, the number of generated electrons significantly increases as the dynamically assisted Schwinger mechanism~\cite{
schutzhold2008dynamically}.

\section*{Acknowledgement}
We thank H. Nakazato for helpful discussions. T.S. was supported by Japan's MEXT Quantum Leap Flagship Program Grant No. JPMXS0120319794.

\appendix

\section{A perturbative derivation of the transition probability}\label{sec:append_1st}
We derive the approximate solution~\eqref{eq:approx_1st} from Eq.~\eqref{eq:approx_1st_numerical}. $f$ and $g$ are components of the time evolution operator $U_{\text{LZ}}$. These are obtained analytically and are expressed as linear combinations of parabolic cylinder functions as Eq.~\eqref{eq:unitary}. By using Eq.~\eqref{eq:approx_1st_numerical}, the transition probability is expressed up to first order in $\lambda$ as
\begin{widetext}
\begin{align}
\ev{\overline{P(\tau_f)}}=&|f(\tau_f,\tau_i)|^2-4\lambda\sqrt{\kappa}(|f(\tau_f,\tau_i)|^2-|g(\tau_f,\tau_i)|^2)\int^{\tau_f}_{\tau_i}d\tau\ |f(\tau,\tau_i)|^2|g(\tau,\tau_i)|^2  \\
&+4\lambda\sqrt{\kappa} \operatorname{Re}\qty(f(\tau_f,\tau_i)g(\tau_f,\tau_i)\int^{\infty}_{-\infty}d\tau\ f^*(\tau,\tau_i)g^*(\tau,\tau_i)(|f(\tau,\tau_i)|^2-|g(\tau,\tau_i)|^2)).
\end{align}
\end{widetext}
Using the equations for a positive value of $\tau$
\begin{align}
D_{i\kappa}\qty(e^{\frac{3\pi}{4}i}\tau)&= e^{\frac{i}{4}\tau^2}e^{-\frac{3\pi}{4}\kappa}\tau^{i\kappa}+\mathcal{O}(\tau^{-1}),
\end{align}
\begin{align}
D_{-i\kappa}\qty(e^{\frac{\pi}{4}i}\tau)&= e^{-\frac{i}{4}\tau^2}e^{\frac{\pi}{4}\kappa}\tau^{-i\kappa}+\mathcal{O}(\tau^{-1}),
\end{align}
\begin{align}
D_{-i\kappa-1}\qty(e^{\frac{5\pi}{4}i}\tau)&= \frac{\sqrt{2 \pi}}{\Gamma(i\kappa+1)}  e^{\frac{i}{4}\tau^{2}}e^{-\frac{\pi}{4}\kappa}\tau^{i\kappa}+\mathcal{O}(\tau^{-1}),
\end{align}
\begin{align}
D_{i\kappa-1}\qty(e^{-\frac{\pi}{4}i}\tau)&= \mathcal{O}(\tau^{-1}),
\end{align}
$|\Gamma(1-ix)|^2=\frac{\pi x}{\sinh(\pi x)}$ and limiting $\tau_i\to-\infty$, $\tau_f\to\infty$, the transition probability is obtained as
\begin{widetext}
\begin{align}
\ev{\overline{P(\infty)}}=e^{-2\pi\kappa}+4\lambda\sqrt{\kappa}\int^{\infty}_{-\infty}d\tau\ \qty(\frac{1-2e^{-2\pi\kappa}}{1-e^{-2\pi\kappa}}e^{2\pi\kappa}|\tilde X_{\kappa}(\tau)|^2+\operatorname{Re}(\tilde X_{\kappa}(\tau)) \tilde Y_{\kappa}(\tau)),
\end{align}
\begin{align}
\tilde X_{\kappa}(\tau)=\frac{\sqrt{2\pi\kappa}}{\Gamma(i\kappa+1)} e^{-2\pi\kappa} D_{i\kappa}(e^{-\frac{\pi}{4}i}\tau)\sqrt{\kappa}D_{i\kappa-1}(e^{-\frac{\pi}{4}i}\tau), \quad
\tilde Y_{\kappa}(\tau)=e^{-\frac{\pi\kappa}{2}}\qty(|D_{i\kappa}(e^{-\frac{\pi}{4}i}\tau)|^2-|\sqrt{\kappa}D_{i\kappa-1}(e^{-\frac{\pi}{4}i}\tau)|^2).
\end{align}
\end{widetext}
Using the formula
\begin{align}
\frac{\sqrt{2\pi}}{\Gamma(\nu+1)}D_{\nu}(z)=e^{-\frac{\pi\nu}{2}i}D_{-\nu-1}(-iz)+e^{\frac{\pi\nu}{2}i}D_{-\nu-1}(iz),
\end{align}
we obtain Eq.~\eqref{eq:approx_1st}. 

\section{The identification of the dominant term}\label{sec:dominant}
\begin{figure}[H]
\centering
\includegraphics[width=83mm]{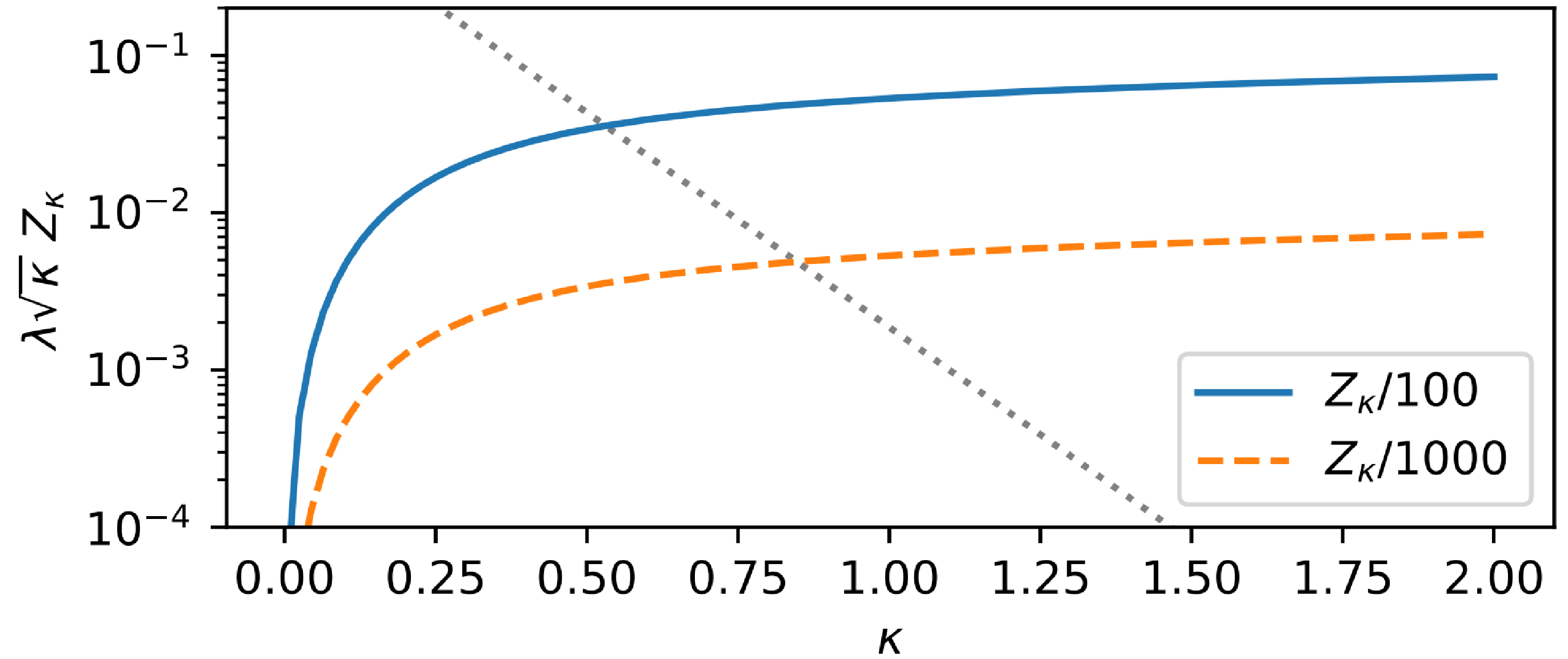}
\caption{Plots of $Z_{\kappa}/100$ (blue), and $Z_{\kappa}/1000$ (orange) against $\kappa$. The gray dotted line represents $\exp\qty(-2\pi\kappa)$. As $\kappa$ increases, it is observed that $Z_{\kappa}$ increases. Eq.~\eqref{eq:approx(i)} is effective if the product of $Z_{\kappa}$ and $\lambda\sqrt{\kappa}$ is significantly smaller than $\exp\qty(-2\pi\kappa)$.}
\label{figure/dominant_small}
\end{figure}

We determine the dominant term in Eq.~\eqref{eq:approx_1st}. First, consider the case of small $\kappa$. In this region, the first term of Eq.~\eqref{eq:approx_1st} is dominant because $\lambda$ is sufficiently small, which leads to the approximate solution~\eqref{eq:approx(i)}. We numerically determine how small $\lambda$ needs to be to dominate the first term. We represent Eq.~\eqref{eq:approx_1st} as
\begin{align}
\ev{\overline{P(\infty)}} \simeq e^{-2\pi\kappa}+\lambda \sqrt{\kappa} Z_{\kappa}.
\end{align}
If $\lambda \sqrt{\kappa} Z_{\kappa}$ is much smaller than $\exp\qty(-2\pi\kappa)$, the dominant term is the first term. Here, $Z_{\kappa}$ is given by
\begin{widetext}
\begin{align}
Z_{\kappa}=
4\int^{\infty}_{-\infty}d\tau\ \qty(|X_{\kappa}(\tau)|^2+\frac{2e^{-2\pi\kappa}}{1-e^{-2\pi\kappa}}\qty(\operatorname{Re}(X_{\kappa}(\tau)))^2+\frac{2e^{-\pi\kappa}}{1-e^{-2\pi\kappa}}\operatorname{Re}\qty(X_{\kappa}(\tau)) Y_{\kappa}(\tau)).
\end{align}
\end{widetext}
Figure~\ref{figure/dominant_small} shows the plots of $\exp\qty(-2\pi\kappa)$, $Z_{\kappa}/100$, and $Z_{\kappa}/1000$. $Z_{\kappa}/100$ corresponds to the case where $\lambda\sqrt{\kappa}=0.01$, while $Z_{\kappa}/1000$ corresponds to $\lambda\sqrt{\kappa}=0.001$. To explore other values of $\lambda\sqrt{\kappa}$, one can simply vertically shift the log plot of $Z_{\kappa}/100$. According to Fig.~\ref{figure/dominant_small}, for example, if $\lambda\sqrt{\kappa}=0.01$, $\exp\qty(-2\pi\kappa)$ dominates when $\kappa\ll0.53$. If $\lambda\sqrt{\kappa}=0.001$, $\exp\qty(-2\pi\kappa)$ dominates when $\kappa\ll0.84$.

\begin{figure}[H]
\centering
\includegraphics[width=83mm]{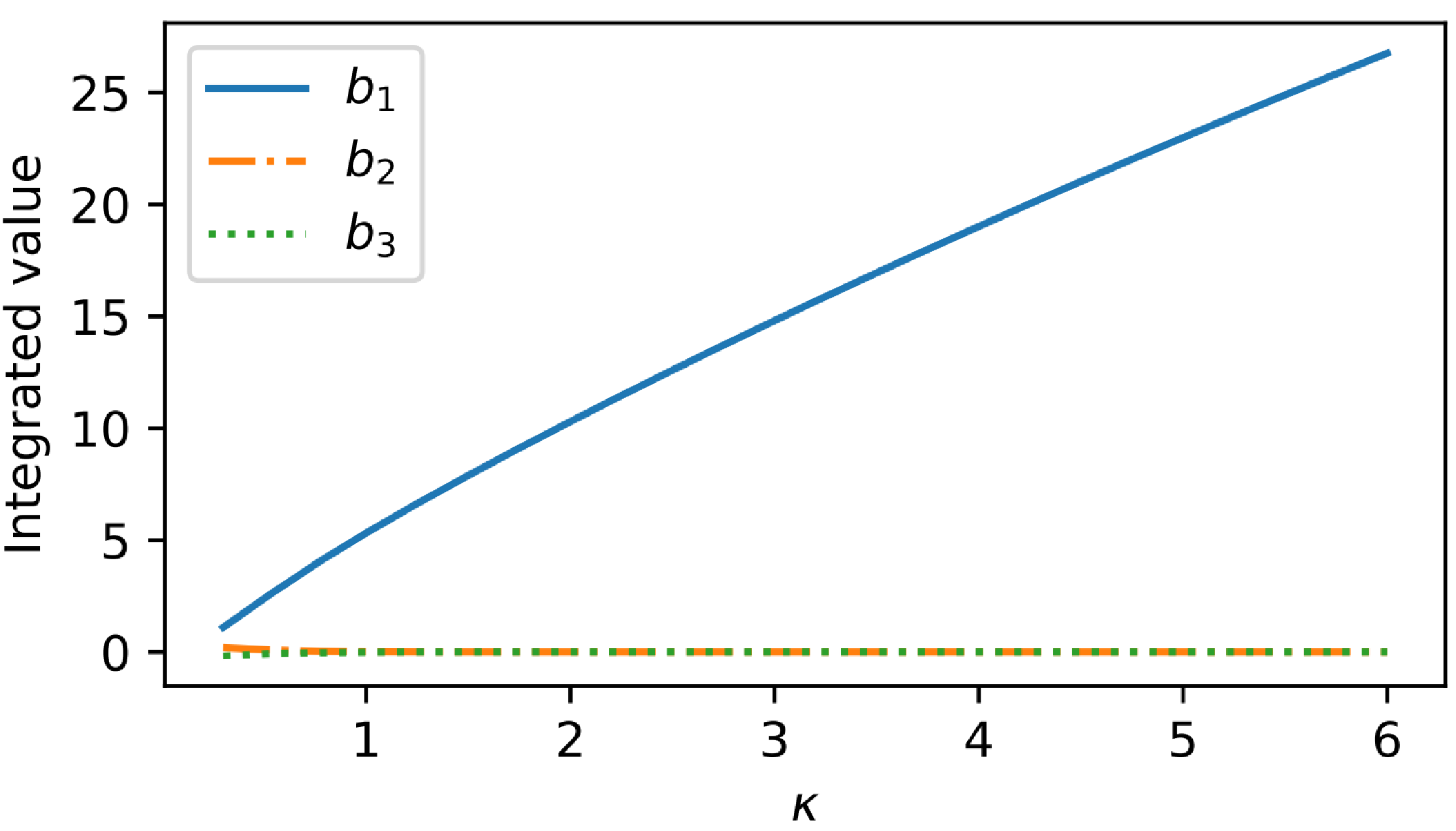}
\caption{Comparison of the three terms proportional to $\lambda$ in Eq.\eqref{eq:approx_1st}. $b_1$ is the largest, indicating its dominance.}
\label{figure/dominant_large}
\end{figure}

Second, in the large $\kappa$ region, we numerically confirm that the term with $\qty|X_{\kappa}|^2$ is dominant. Since $\kappa\gg1$, $e^{-2\pi\kappa}\ll1$ holds. Therefore, it is sufficient to consider only the three terms proportional to 
$\lambda$ in Eq.~\eqref{eq:approx_1st}. The three terms are given by 
\begin{align}
b_1(\kappa)&=4\sqrt{\kappa}\int^{\infty}_{-\infty}d\tau\ |X_{\kappa}(\tau)|^2,\\
b_2(\kappa)&=4\sqrt{\kappa}\frac{2e^{-2\pi\kappa}}{1-e^{-2\pi\kappa}}\int^{\infty}_{-\infty}d\tau\ \qty(\operatorname{Re}(X_{\kappa}(\tau)))^2,\\
b_3(\kappa)&=4\sqrt{\kappa}\frac{2e^{-\pi\kappa}}{1-e^{-2\pi\kappa}}\int^{\infty}_{-\infty}d\tau\ \operatorname{Re}\qty(X_{\kappa}(\tau)) Y_{\kappa}(\tau).
\end{align}
$b_1,\ b_2,\ b_3$ are plotted in Fig.~\ref{figure/dominant_large}. It is observed that $b_1$ is dominant, which leads Eq.~\eqref{eq:b1}.

\section{Calculation of the integral of a product of special functions}\label{sec:append_approx1}
We demonstrate Eq.~\eqref{eq:approx(ii)} in the region (ii) ($1\ll \kappa\ll \frac{1}{\lambda}$). The approximate solution for the transition probability is approximated as
\begin{align}
\ev{\overline{P(\infty)}}&\simeq 4\lambda\sqrt{\kappa}e^{-\pi\kappa}\kappa^2\\
&\quad \times\int^{\infty}_{-\infty}d\tau\ \qty|D_{-i\kappa-1}(e^{\frac{\pi}{4}i}\tau)D_{-i\kappa-1}(-e^{\frac{\pi}{4}i}\tau)|^2.
\end{align}
To perform this integration, we utilize the following formula~\cite{nasri2016product}
\begin{align}
&D_{-\nu}(z)D_{-\nu}(-z)\\
=&\frac{2e^{\frac{z^2}{2}}}{\Gamma\qty(\nu+\frac{1}{2})} \int^{\infty}_0dy\ \cos (2zy)e^{-2y^2} {}_1F_1\qty(\frac{1}{2},\nu+\frac{1}{2},y^2).
\end{align}
Here ${}_1F_1$ denotes the confluent hypergeometric function~\cite{gradshteyn2014table}. By substituting $z=e^{i\frac{\pi}{4}}\tau$ and $\nu=i\kappa+1$, and changing the integration path to $r=e^{i\frac{\pi}{4}}y$, we get
\begin{align}
&D_{-i\kappa-1}(e^{i\frac{\pi}{4}}\tau)D_{-i\kappa-1}(-e^{i\frac{\pi}{4}}\tau)\\
=&\frac{2e^{-i\frac{\pi}{4}}e^{i\frac{\tau^2}{2}}}{\Gamma\qty(i\kappa+\frac{3}{2})} \int^{\infty}_0dr\ \cos (2\tau r)e^{2ir^2} {}_1F_1\qty(\frac{1}{2},i\kappa+\frac{3}{2},-ir^2).
\end{align}
Therefore, $\ev{\overline{P(\infty)}}$ is approximated as
\begin{widetext}
\begin{align}
\ev{\overline{P(\infty)}}
&\simeq 4\lambda\sqrt{\kappa}e^{-\pi\kappa}\kappa^2\int^{\infty}_{-\infty}d\tau\ \frac{1}{\qty|\Gamma\qty(i\kappa+\frac{3}{2})|^2} \int^{\infty}_0dr\ (e^{2i\tau r}+e^{-2i\tau r})e^{2ir^2} {}_1F_1\qty(\frac{1}{2},i\kappa+\frac{3}{2},-ir^2) \\
&\quad \times \int^{\infty}_0ds\ (e^{2i\tau s}+e^{-2i\tau s})e^{-2is^2} {}_1F_1\qty(\frac{1}{2},-i\kappa+\frac{3}{2},is^2).
\end{align}
\end{widetext}
We change the order of integration and perform the integration over $\tau$ first. Given that $r$ and $s$ are positive, we can employ the formulas
\begin{align}
\int^{\infty}_{-\infty}d\tau\ e^{2i\tau(r-s)}=\pi\delta(r-s),\ 
\int^{\infty}_{-\infty}d\tau\ e^{\pm2i\tau(r+s)}=0.
\end{align}
Consequently, we get
\begin{align}
\ev{\overline{P(\infty)}}\simeq&4\lambda\sqrt{\kappa}\frac{2\pi}{\qty|\Gamma\qty(i\kappa+\frac{3}{2})|^2}e^{-\pi\kappa}\kappa^2\\ &\times\int^{\infty}_0dr\ \qty|{}_1F_1\qty(\frac{1}{2},i\kappa+\frac{3}{2},-ir^2)|^2.
\end{align}
Note that this derivation is based on the assumption of infinite time integration. Therefore, this technique cannot be applied in systems where the time span is finite. Using the following formula
\begin{align}
\frac{2\pi}{\qty|\Gamma\qty(i\kappa+\frac{3}{2})|^2}e^{-\pi\kappa}\kappa^2=\frac{\kappa^2}{\kappa^2+\frac{1}{4}}\qty(1+e^{-2\pi\kappa})
\end{align}
and the approximations $e^{-2\pi\kappa}\ll 1$ and $\kappa^2\gg1/4$, the transition probability is expressed as
\begin{align}
\ev{\overline{P(\infty)}}\simeq4\lambda\sqrt{\kappa}\int^{\infty}_0dr\ \qty|{}_1F_1\qty(\frac{1}{2},i\kappa+\frac{3}{2},-ir^2)|^2.
\end{align}
By employing the variable transformation $r=\sqrt{\kappa}s$, the expression can be
\begin{align}
\ev{\overline{P(\infty)}}\simeq4\lambda\kappa\int^{\infty}_0ds\ \qty|{}_1F_1\qty(\frac{1}{2},i\kappa+\frac{3}{2},-i\kappa s^2)|^2.
\end{align}
Furthermore, when $\kappa$ is large, the following equation holds:
\begin{align}
{}_1F_1\qty(\frac{1}{2},i\kappa+\frac{3}{2},-i\kappa s^2)= \frac{1}{\sqrt{1+s^2}}+\mathcal{O}\qty(\kappa^{-1})
\label{eq:1F1_approx}
\end{align}
We confirm in Appendix \ref{sec:append_approx2} that Eq.~\eqref{eq:1F1_approx} is effective. Consequently, by using
\begin{align}
4\lambda\kappa\int^{\infty}_0\frac{ds}{1+s^2}=2\pi\lambda\kappa,
\end{align}
we have demonstrated the validity of Eq.~\eqref{eq:approx(ii)}.

\section{The asymptotic behavior of the special function}\label{sec:append_approx2}
\begin{figure}[H]
\centering
\includegraphics[width=85mm]{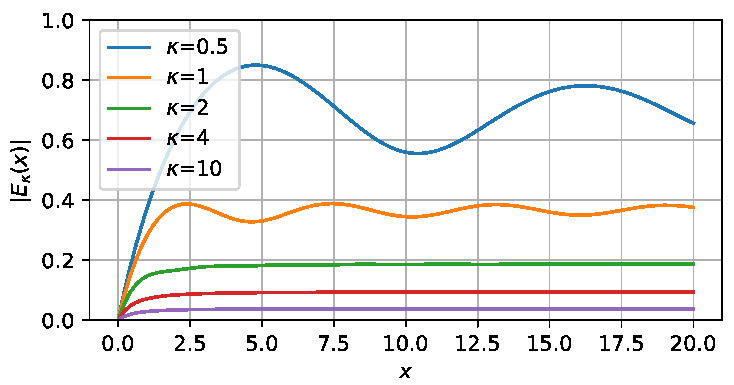}
\caption{Plots of $\qty|E_{\kappa}(x)|$ against $x$. As $\kappa$ increases, it is observed that this value becomes significantly smaller than 1 for any $x$.}
\label{figure/compare_with_1}
\end{figure}

As stated in Appendix \ref{sec:append_approx1}, the approximate formula~\eqref{eq:1F1_approx} is shown to be effective when $\kappa$ is large. With the properties of a confluent hypergeometric function~\cite{gradshteyn2014table}
\begin{align}
\frac{d}{dz}{}_1F_1\qty(a,b,z)=\frac{a}{b}{}_1F_1\qty(a+1,b+1,z)
\end{align}
and 
\begin{align}
{}_1F_1\qty(a,b,z)=&\frac{b-a-z-1}{b}{}_1F_1\qty(a+1,b+1,z)\\
&+\frac{a+1}{b}{}_1F_1\qty(a+2,b+1,z),
\end{align}
we get
\begin{align}
&\frac{d}{dx}\qty(\sqrt{1+x}{}_1F_1\qty(\frac{1}{2},i\kappa+\frac{3}{2},-i\kappa x)) \\
=&\frac{3}{4\sqrt{1+x}(i\kappa+\frac{3}{2})}{}_1F_1\qty(\frac{5}{2},i\kappa+\frac{5}{2},-i\kappa x).
\end{align}
Integrating both sides with $x$ from 0 and using ${}_1F_1\qty(a, b, 0) = 1$ yield
\begin{align}
{}_1F_1\qty(\frac{1}{2},i\kappa+\frac{3}{2},-i\kappa x)=\frac{1}{\sqrt{1+x}}\qty(1+E_{\kappa}(x)),
\end{align}
\begin{align}
E_{\kappa}(x)=\frac{3}{4(i\kappa+\frac{3}{2})}\int^x_0\frac{dy}{\sqrt{1+y}}{}_1F_1\qty(\frac{5}{2},i\kappa+\frac{5}{2},-i\kappa y).
\end{align}
Therefore, Eq.~\eqref{eq:1F1_approx} is valid if $E(x)\ll 1$. $\qty|E_{\kappa}(x)|$ against $x$ are plotted in Fig.~\ref{figure/compare_with_1}. It can be observed $\qty|E_{\kappa}(x)|\ll1$ for large $\kappa$, confirming the validity of the approximation Eq.~\eqref{eq:1F1_approx}.

\section{The validity of perturbations in the adiabatic approximation}\label{sec:append_adiabatic}
We discuss why the approximate solution~\eqref{eq:approx(ii)(iii)}, which is obtained using the adiabatic approximation, remains valid to all orders in $\lambda$ when $\kappa$ is large. According to the adiabatic approximation, as demonstrated in Eq.~\eqref{eq:integration_tau}, it is necessary to perform the integration
\begin{align}
&\quad c_{2}(\tau)\\
&\simeq c_{2}(\tau_i) \exp\qty(\int^{\tau}_{\tau_i}d\tau' \chi_{2}(\tau'))\\
&= c_{2}(\tau_i) \exp\qty(-\int^{\tau}_{\tau_i}d\tau' \sqrt{\kappa}\qty(\frac{2\lambda}{z^2(\tau')+1}+\mathcal{O}(\lambda^3))).
\end{align}
As shown in Eq.~\eqref{eq:integration_Z}, the exponent is multiplied by $\sqrt{\kappa}$ due to the Jacobian of the integration with respect to $z$. Considering that $\chi_2$ is proportional to $\sqrt{\kappa}$, all terms in the exponent are proportional to $\kappa$. By taking $\tau_i\to-\infty$ and $\tau\to\infty$, we have
\begin{align}
c_{2}(\infty)&\simeq 
c_{2}(-\infty) \exp\qty(-\int^{\infty}_{-\infty}dz\ 2\kappa\qty(\frac{2\lambda}{{z}^2+1}+\mathcal{O}(\lambda^3)))\\
&=c_{2}(-\infty) \exp\qty(-4\pi\lambda\kappa) \exp\qty(\int^{\infty}_{-\infty}dz\ \kappa\mathcal{O}(\lambda^3)).
\end{align}
Although the integrand 
$\mathcal{O}(\lambda^3)$ depends on $z$, we have
\begin{align}
\chi_2(z)=-\sqrt{\kappa}\qty(\frac{2\lambda}{{z}^2}-\frac{2(\lambda+\lambda^3)}{{z}^4}+\mathcal{O}\qty({z}^{-6})),
\end{align}
where the powers of $z$ in $\mathcal{O}(\lambda^3)$ are less than $-2$. As a result, the integration of $\mathcal{O}(\lambda^3)$ over $z$ contributes negligibly. Therefore, $c_{2}(\infty)$ can be approximated as
\begin{align}
c_{2}(\infty)&\simeq c_{2}(-\infty) \exp\qty(-4\pi\lambda\kappa) \exp\qty(\kappa\mathcal{O}\qty(\lambda^3)).
\end{align}
Considering the terms proportional to $\lambda^m$ in $c_{2}(\infty)$, the factor $\exp\qty(-4\pi\lambda\kappa)$ contributes with a power of $\kappa$ equal to $m$. In contrast, the neglected factor $\exp\qty(\kappa\mathcal{O}\qty(\lambda^3))$ has a lower power of $\kappa$ than $m$, leading to a minor contribution. Thus, for large $\kappa$, the approximation solution~\eqref{eq:approx(ii)(iii)} remains valid to all orders of $\lambda$.

\section{The comparisons with the Kayanuma formula}\label{sec:append_Kayanuma}
We first compute the transition probability of noise-induced Landau--Zener model in the regime where $\kappa$ is sufficiently large ($\lambda\kappa\gg1$). Next, we verify the defect density within the transverse field Ising model under this condition. Our approximate solutions of the transition probability and the defect density are compared with those derived from the Kayanuma formula, which is effective in large $\lambda$.

\begin{figure}[H]
\centering
\includegraphics[width=85mm]{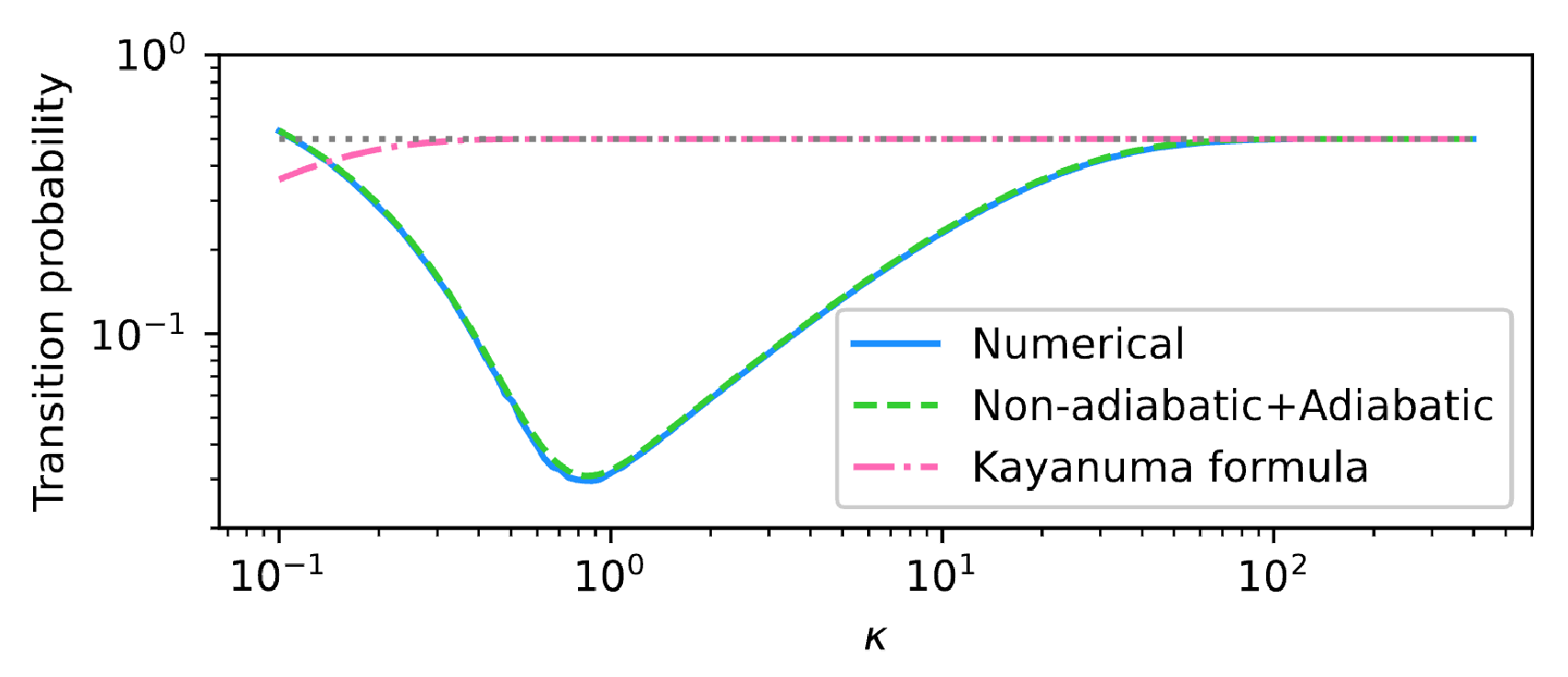}
\caption{Plots of the transition probabilities at $\tau_f$ against $\kappa$. The blue line (Numerical) represents the numerical simulation of Eq.~\eqref{eq:mastereq_int} over the time span from $\tau_i=-200$ to $\tau_f=200$. The green dashed line (Non-adiabatic+Adiabatic) represents $p^{\text{non-ad}}+p^{\text{ad}}$, while the red dashed line (Kayanuma formula) represents $p^{\text{Kayanuma}}$, with $\lambda=10^{-3}$. The gray dotted line represents $1/2$.}
\label{figure/kayanuma1}
\end{figure}

First, we verify the behavior of the transition probability in the regime where $\kappa$ is sufficiently large. The approximate solution~\eqref{eq:approx(i)(ii)(iii)} approaches $1/2$ as $\kappa\to\infty$, similar to the Kayanuma formula~\cite{kayanuma1984nonadiabatic}
\begin{align}
\ev{\overline{P(\infty)}}&\simeq \frac{1}{2}\qty(1- \exp\qty(-4\pi\kappa))\\
&=:p^{\text{Kayanuma}}(\kappa).
\label{eq:Kayanuma1}
\end{align}
The comparison between $p^{\text{non-ad}}+p^{\text{ad}}$ and $p^{\text{Kayanuma}}$ is shown in Fig.~\ref{figure/kayanuma1}. In the regime where $\kappa$ is large, all cases asymptotically approach $1/2$. However, in the small $\kappa$ region, $p^{\text{Kayanuma}}$ fails to reproduce the behavior of the numerical results.

\begin{figure}[H]
\centering
\includegraphics[width=85mm]{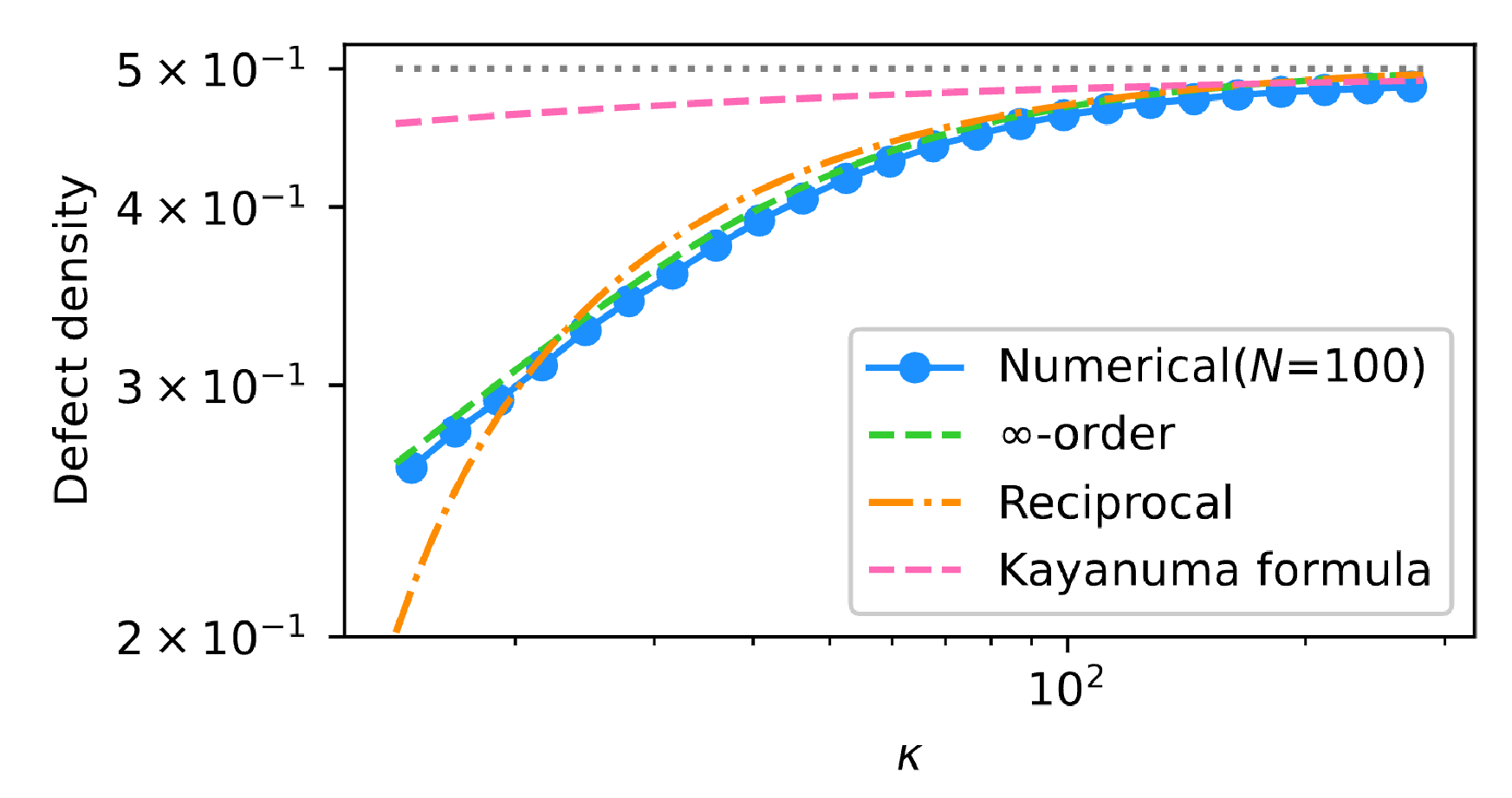}
\caption{Plots of the defect density against $\kappa$, focusing around $\kappa\gg1$. The blue solid line (Numerical) represents the numerical simulation of Eq.~\eqref{eq:mastereq_int2} over the time span $\tau_i=-200$ to $\tau_f=200$, plotting the discrete sum in Eq.~\eqref{eq:defect}, with $N=100$. The green dashed line ($\infty$-order) represents $n^{\infty\text{-order}}$, the yellow dashed line (Reciprocal) represents $n^{\text{reciprocal}}$, and the pink dashed line (Kayanuma) represents $n^{\text{Kayanuma}}$, with $\lambda=5\times 10^{-3}$. The gray dotted line represents $1/2$.}
\label{figure/kayanuma2}
\end{figure}

Next, we verify the defect density in the large $\kappa$ region. In a previous study~\cite{dutta2016anti}, $n$ is obtained by integrating Eq.~\eqref{eq:Kayanuma1} with respect to $q$ as
\begin{align}
n^{\text{Kayanuma}}(\kappa) \simeq \frac{1}{2} - \frac{1}{4\pi \sqrt{\kappa}}.
\end{align}
On the other hand, by employing the asymptotic formulas of the functions $I_0(z)$ and $L_0(z)$ to Eq.~\eqref{eq:defect_infty}, $n$ is approximated by neglecting $\mathcal{O}(4\pi\lambda\kappa)^{-2}$ as
\begin{align}
n^{\infty\text{-order}}(\lambda,\kappa)&\simeq
n^{\text{KZM}}(\kappa)+ \frac{1}{2}-\frac{1}{4\pi^2\lambda\kappa}\\
&=: n^{\text{reciprocal}}(\lambda,\kappa).
\end{align}
We plot $n$ obtained numerically, along with $n^{\infty\text{-order}}$, $n^{\text{reciprocal}}$, and $n^{\text{Kayanuma}}$ in Fig.~\ref{figure/kayanuma2}. We can confirm that $n^{\text{reciprocal}}$ is closer to numerical results than $n^{\text{Kayanuma}}$.

\bibliographystyle{ytphys.bst}
\bibliography{ref} 
\end{document}